\documentclass[11pt,a4paper]{article}

 \usepackage{amssymb,amsfonts,amsmath,amsthm,mathrsfs}
 \usepackage[english]{babel}
 \usepackage[utf8]{inputenc}  
 \usepackage{enumitem}
 \usepackage{url}
 \usepackage{graphicx}
 \usepackage{color}
 \usepackage[font=small,labelfont=bf]{caption} 
 \usepackage{empheq,tikz}
 \usepackage{vmargin}
 \usepackage{minitoc}
 \usepackage[a4paper, margin=1in]{geometry}
 \usepackage{pdfpages} 
\usepackage{afterpage}
\usepackage{bm}
\usepackage[toc,page]{appendix}
\usepackage{multicol}

\usepackage[textsize=normalsize,textwidth=2cm,colorinlistoftodos]{todonotes}
\newcounter{todocounter}
\newcommand{\todonum}[2][]
{\stepcounter{todocounter}\todo[#1]{#2}}

\newcommand{\kevish}[1]{\todonum[inline,color=blue!20,textcolor=black]{{\bf Kevish (\thetodocounter):} #1}}
\newcommand{\art}[1]{\todonum[inline,color=green!40,textcolor=black]{{\bf Art (\thetodocounter):} #1 }}

\usepackage[most]{tcolorbox} 

\newtcolorbox[auto counter]{optionalnote}[2][]{
    parbox=false,
    colbacktitle= white,
    colback=green!5!white,
    colframe=white!45!black,
    coltitle=black,
    enhanced,
    attach boxed title to top left={yshift=-1mm},
    title={\thetcbcounter.~#2}
,#1}

\newtcolorbox{highlight-result}[1][]{
 parbox=false,
 boxrule=0pt,top=0pt,bottom=0pt,
colback=blue!8!white,
enhanced,#1}

\tcbset{colback=blue!8!white, colframe=white!50!white,top=0mm,bottom=0mm,right=0mm,left=0mm}

\usepackage{fancyhdr}
 \usepackage[yyyymmdd,hhmmss]{datetime}
 \AtEndDocument{\label{lastpage}}
\setmargrb
 {2.5cm} 
 {1.5cm} 
 {2.5cm} 
 {2cm} 
 
 \usepackage{hyperref}
\hypersetup{
pdfpagemode=UseNone,
pdftoolbar=true, 
pdfmenubar=true, 
pdffitwindow=false, 
pdfstartview={Fit}, 
pdftitle={}, 
pdfauthor={}, 
pdfsubject={}, 
pdfcreator={}, 
pdfproducer={}, 
pdfkeywords={}, 
pdfnewwindow=true, 
colorlinks=true, 
linkcolor=magenta, 
citecolor=red, 
filecolor=cyan, 
urlcolor=blue 
}

 \usepackage{cleveref}

\author{K. Napal, P. Piva, A. L. Gower}

\usetikzlibrary{mindmap,trees,shadows,shapes}

\definecolor{bronze}{rgb}{0.8, 0.5, 0.2}
\definecolor{coralred}{rgb}{1.0, 0.25, 0.25}
\definecolor{coralpink}{rgb}{0.97, 0.51, 0.47}
\definecolor{lightpink}{rgb}{1.0, 0.71, 0.76}
\definecolor{lightskyblue}{rgb}{0.53, 0.81, 0.98}
\definecolor{applegreen}{rgb}{0.55, 0.71, 0.0}
\definecolor{Darm}{RGB}{51,51,178}
\definecolor{saffron}{rgb}{0.96, 0.77, 0.19}

\newcommand{\DrawCylinder}[3]
{
\filldraw[fill=bronze!90,draw=black] (\ox,\h/2 + \oy) ellipse (\a cm and \a*\ellratio cm);

\fill[fill=bronze,opacity=.6,dashed]
(-\a + \ox,\oy) 
 -- (-\a + \ox,\h/2 + \oy) 
-- plot [domain=pi:2*pi] ({\a*cos(\x r)+\ox},{\a*\ellratio*sin(\x r)+\h/2 + \oy})
-- plot [domain=2*pi:pi] ({\a*cos(\x r) + \ox},{\a*\ellratio*sin(\x r) + \oy}) 
 -- cycle;
\draw (\ox-\a,\oy) -- (\ox-\a,\oy + \h/2);
\draw (\ox+\a,\oy) -- (\ox+\a,\oy + \h/2);

\filldraw[fill=bronze,draw=bronze,opacity=.3,dashed]
(-\a + \ox,-\h/2 + \oy) 
 -- (-\a + \ox, \oy) 
-- plot [domain=pi:2*pi] ({\a*cos(\x r)+\ox},{\a*\ellratio*sin(\x r)+ \oy})
-- plot [domain=2*pi:pi] ({\a*cos(\x r) + \ox},{\a*\ellratio*sin(\x r)-\h/2 + \oy}) 
 -- cycle;

\draw[dashed,draw=black] (\ox,\oy) ellipse (\a cm and \a*\ellratio cm);

\draw (\ox,\oy) node[below right] {$\mathbf{r}_{#1}$} node{$\bullet$};
}

\newcommand{\TikzCylinders}[1]
{
\begin{tikzpicture}[scale=#1]

\def\a{1.2};
\def\h{3};
\def\R{7}
\def\Rtilde{5.8}
\def\ellratio{.5};

\filldraw[dashed,gray!10] (0,0) ellipse (\R cm and \ellratio*\R cm);

\def\RR{1.2*\R};
\def\Rshift{1};
\fill[gray!40] (-\RR-2*\Rshift,-\RR*\ellratio) -- 
(-\RR+\Rshift,\RR*\ellratio) -- (\RR+\Rshift,\RR*\ellratio)
 -- (\RR-\Rshift,-\RR*\ellratio) -- cycle;

\draw[dashed] (0,0) ellipse (\R cm and \ellratio*\R cm);
    \filldraw[dashed,draw=applegreen,fill=applegreen!50] (0,0) ellipse (\Rtilde cm and \ellratio*\Rtilde cm);

\foreach \i/\ox/\oy in
	{	
		 2/-.6*\Rtilde/.3*\Rtilde,
		 1/-.4*\Rtilde/-.3*\Rtilde, 
		 3/-.1*\Rtilde/.48*\Rtilde, 
		 4/.7*\Rtilde/.3*\Rtilde
	}
	{
		\DrawCylinder{\i}{\ox}{\oy}
	}

\draw (0,0) -- (.7*6.3,-.3*6.3) node[midway,below,sloped] {$\tilde{R}=R-a_1$};
\draw (0,0) -- (\R,0) node[midway,below,sloped] {$R$};

\draw[applegreen] (0,-.8*\ellratio*\Rtilde) node[scale=1.5]{$\mathcal{R}_1$};
  
\draw (0,0) node[above left] {$O$} node{$\bullet$};
 \draw (-.4*\Rtilde,-.3*\Rtilde) -- (-.4*\Rtilde+\a,-.3*\Rtilde) node[midway,above,sloped] {$a_1$};

 \coordinate (O) at (-8,-2);
 \coordinate (A) at (-8-1*.7,-2-1.9*.7);
  \coordinate (B) at (-8,-2+2);
 \coordinate (C) at (-8+2,-2);

  \draw[-latex] (O) -- (A) node[below] {$x$};
  \draw[-latex] (O) -- (B) node[above] {$z$};
  \draw[-latex] (O) -- (C) node[below] {$y$};

\end{tikzpicture}
}

\tikzset{>=latex} 
\colorlet{myred}{red!85!black}
\colorlet{myblue}{orange!80!black}
\colorlet{mydarkred}{myred!80!black}
\colorlet{mydarkblue}{orange!80!black}
\tikzstyle{xline}=[myblue,very thick]

\title{Effective T-matrix of a cylinder filled with a random 2D particulate}

\begin{document} 

\newcommand{\dsp}{\displaystyle}


\newcommand{\edit}[1]{{\color{red}#1}}

\newcommand{\sphere}{\ensuremath{{\mathbb S^2}}}
\newcommand{\dsphere}{\ensuremath{\mathbb S^{d-1}}}
\newcommand{\tsphere}{\ensuremath{\mathbb S^{2}}}

\newcommand{\A}{\ensuremath{\mathbb A}}
\newcommand{\B}{\ensuremath{\mathbb B}}
\newcommand{\C}{\ensuremath{\mathbb C}}
\newcommand{\E}{\ensuremath{\mathbb E}}
\newcommand{\K}{\ensuremath{\mathbb K}}
\newcommand{\Q}{\ensuremath{\mathbb Q}}
\newcommand{\R}{\ensuremath{\mathbb R}}
\newcommand{\T}{\ensuremath{\mathbb T}}
\newcommand{\U}{\ensuremath{\mathbb U}}
\newcommand{\V}{\ensuremath{\mathbb V}}
\newcommand{\W}{\ensuremath{\mathbb W}}
\newcommand{\X}{\ensuremath{\mathbb X}}
\newcommand{\Y}{\ensuremath{\mathbb Y}}
\newcommand{\Z}{\ensuremath{\mathbb Z}}

\newcommand{\Gammabar}{\ensuremath{{\overline{\Gamma}}}}
\newcommand{\Omegabar}{\ensuremath{{\overline{\Omega}}}}

\newcommand{\gammax}{\ensuremath{{\d D\setminus \Omega}}}
\newcommand{\gammay}{\ensuremath{{\d \Omega \cap D}}}
\newcommand{\gammaz}{\ensuremath{{\d \Omega \setminus D}}}
\newcommand{\LXL}{\ensuremath{{L^2(D\setminus \overline{\Omega})\times L^2(\d \Omega)}}}
\newcommand{\LXLG}{\ensuremath{{L^2(D\setminus \Gamma)\times L^2(\Gamma)}}}
\newcommand{\LXLO}{\ensuremath{{L^2(D\setminus \Omega)\times L^2(\Omega)}}}
\newcommand{\DSO}{\ensuremath{{D\setminus\d\Omega}}}

\newcommand{\Leb}[1]{\ensure\documentclass[11pt]{article}
 \usepackage{amssymb,amsfonts,amsmath,amsthm,mathrsfs}
 \usepackage[english]{babel}
 \usepackage[utf8]{inputenc}  
 \usepackage{enumitem}
 \usepackage{url}
 \usepackage{graphicx}
 \usepackage{color}
 \usepackage[font=small,labelfont=bf]{caption} 
 \usepackage{empheq,tikz}
 \usepackage{vmargin}
 \usepackage{minitoc}
 \usepackage[a4paper, margin=1in]{geometry}
 \usepackage{pdfpages} 
\usepackage{afterpage}
\usepackage{bm}

\usepackage[textsize=normalsize,textwidth=2cm,colorinlistoftodos]{todonotes}

\newcommand{\kevish}[1]{\todo[inline,color=purple!20]{{\bf Kevish:} #1}}
\newcommand{\art}[1]{\todo[inline,color=green!20]{{\bf Art:} #1}}

\usepackage{fancyhdr}
 \usepackage[yyyymmdd,hhmmss]{datetime}
\pagestyle{fancy}
 \cfoot{\thepage}
 \AtEndDocument{\label{lastpage}}
\lfoot{}
 \lhead{}
 
\setmargrb
 {2.5cm} 
 {1.5cm} 
 {2.5cm} 
 {2cm} 
 

 \usepackage{hyperref}
\hypersetup{
pdfpagemode=none,
pdftoolbar=true, 
pdfmenubar=true, 
pdffitwindow=false, 
pdfstartview={Fit}, 
pdftitle={}, 
pdfauthor={}, 
pdfsubject={}, 
pdfcreator={}, 
pdfproducer={}, 
pdfkeywords={}, 
pdfnewwindow=true, 
colorlinks=true, 
linkcolor=magenta, 
citecolor=red, 
filecolor=cyan, 
urlcolor=blue 
}
 
\hypersetup{
linkcolor=orange, 
citecolor=orange, 
}
\usepackage{xcolor}
\pagecolor[rgb]{0,0,0}
\color[rgb]{1,1,0.5}


math{L^2(#1)}}
\newcommand{\Sob}[2]{\ensuremath{H^{#1}(#2)}}
\newcommand{\Sobfrac}[4]{\ensuremath{{H^{{#1}\frac{#2}{#3}}(#4)}}} 

\def\d{\partial}
\newcommand{\dnu}[1]{\ensuremath{\frac{\partial #1}{\partial \nu}}}
\newcommand{\dr}[1]{\ensuremath{\frac{\partial #1}{\partial r}}}
\newcommand{\vf}{\ensuremath{\varphi}}
\newcommand{\real}{\ensuremath{\Re\mathfrak{e}}}
\newcommand{\image}{\ensuremath{\Im\mathfrak{m}}}

\newcommand{\Norm}[2]{\ensuremath{{\|#1\|}_{#2}}}
\newcommand{\NLeb}[2]{\ensuremath{{\|#1\|}_{L^2(#2)}}}
\newcommand{\NSob}[3]{\ensuremath{{\|#1\|}_{H^{#2}(#3)}}}
\newcommand{\NSobfrac}[5]{\ensuremath{{\|#1\|}_{H^{{#2}\frac{#3}{#4}}(#5)}}}

\newtheorem{theorem}{Theorem}[section]
\newtheorem{lemma}[theorem]{Lemma}
\newtheorem{proposition}[theorem]{Proposition}
\newtheorem{corollary}[theorem]{Corollary}
\newtheorem{definition}[theorem]{Definition}
\newtheorem{hypothesis}[theorem]{Assumption}
\newtheorem{remark}[theorem]{Remark}

\newcommand{\fonction}[5]{\ensuremath{
\begin{array}{rccl}
#1 :& #2 & \longrightarrow & #3
\\
& #4 &\longmapsto & #5
\end{array}
}}

\def \equi#1{\mathrel{\mathop{\kern 0pt\sim}\limits_{#1}}} 


\def\Ac{{\cal A}}
\def\Bc{{\cal B}}
\def\Cc{{\cal C}}
\def\Dc{{\cal D}}
\def\Ec{{\cal E}}
\def\Fc{{\cal F}}
\def\Gc{{\cal G}}
\def\Hc{{\cal H}}
\def\Ic{{\cal I}}
\def\Jc{{\cal J}}
\def\Kc{{\cal K}}
\def\Lc{{\cal L}}
\def\Mc{{\cal M}}
\def\Nc{{\cal N}}
\def\Oc{{\cal O}}
\def\Pc{{\cal P}}
\def\Qc{{\cal Q}}
\def\Rc{{\cal R}}
\def\Sc{{\cal S}}
\def\Tc{{\cal T}}
\def\Uc{{\cal U}}
\def\Vc{{\cal V}}
\def\Wc{{\cal W}}
\def\Xc{{\cal X}}
\def\Yc{{\cal Y}}
\def\Zc{{\cal Z}}

\def\Cfrak{{\mathfrak C}}


\newcommand{\mrB}{\mathrm{B}}
\newcommand{\bfn}{\boldsymbol{n}}
\newcommand{\bfxi}{\boldsymbol{\xi}}
\newcommand{\mB}{\mrm{B}}
\newcommand{\mA}{\mrm{A}}
\newcommand{\spec}{\mathfrak{S}}
\def\cont{\mathscr C}		
\def\csubset{\subset\subset}	
\def\bg{{\mathbb 1}}		
\newcommand{\lbr}{\lbrack}
\newcommand{\rbr}{\rbrack}
\newcommand{\eps}{\varepsilon}
\newcommand{\om}{\omega}
\newcommand{\Om}{\Omega}
\newcommand{\mrm}[1]{\mathrm{#1}}
\newcommand{\freq}{\omega}

\newcommand{\xhat}{\hat{x}}
\newcommand{\dhat}{\hat{d}}
\newcommand{\betahat}{\hat{\beta}}
\newcommand{\mbxhat}{{\boldsymbol{\hat{x}}}}
\newcommand{\mbyhat}{{\boldsymbol{\hat{y}}}}
\newcommand{\mbzhat}{{\boldsymbol{\hat{z}}}}
\newcommand{\mbrhat}{{\boldsymbol{\hat{r}}}}
\newcommand{\mbkhat}{{\mathbf{\hat{k}}}}
\newcommand{\mbtau}{{\mathbf{\tau}}}

\newcommand{\mba}{\boldsymbol{a}}
\newcommand{\mbb}{\boldsymbol{b}}
\newcommand{\mbc}{\boldsymbol{c}}
\newcommand{\mbd}{\boldsymbol{d}}
\newcommand{\mbe}{\boldsymbol{e}}
\newcommand{\mbf}{\boldsymbol{f}}
\newcommand{\mbg}{\boldsymbol{g}}
\newcommand{\mbh}{{\boldsymbol{h}}}
\newcommand{\mbi}{\mathbf{i}}
\newcommand{\mbj}{\mathbf{j}}
\newcommand{\mbk}{\mathbf{k}}
\newcommand{\mbl}{\mathbf{l}}
\newcommand{\mbm}{\boldsymbol{m}}
\newcommand{\mbn}{\boldsymbol{n}}
\newcommand{\mbq}{\boldsymbol{q}}
\newcommand{\mbr}{\boldsymbol{r}}
\newcommand{\mbv}{\mathbf{v}}
\newcommand{\mbw}{\mathbf{w}}
\newcommand{\mbx}{\boldsymbol{x}}
\newcommand{\mby}{\boldsymbol{y}}

\newcommand{\mbt}{\boldsymbol{t}}
\newcommand{\mbu}{\mathbf{u}}

\newcommand{\mbz}{\boldsymbol{z}}

\newcommand{\mbrho}{\boldsymbol{\rho}}
\newcommand{\mbnu}{{\bm{\nu}}}
\newcommand{\mbeps}{\mathbf{\eps}}
\newcommand{\mbdiv}{\mathbf{\mathrm{div}}\,}
\newcommand{\mbnabla}{\mathbf{\mathrm{\nabla}}}
\newcommand{\mbDelta}{\mathbf{\mathrm{\Delta}}}

\newcommand{\mbA}{\mathbf{A}}
\newcommand{\mbB}{\mathbf{B}}
\newcommand{\mbC}{\mathbf{C}}
\newcommand{\mbD}{\mathbf{D}}
\newcommand{\mbE}{\mathbf{E}}
\newcommand{\mbF}{\mathbf{F}}
\newcommand{\mbG}{\mathbf{G}}
\newcommand{\mbH}{\mathbf{H}}
\newcommand{\mbI}{\mathbf{I}}
\newcommand{\mbJ}{\mathbf{J}}
\newcommand{\mbK}{\mathbf{K}}
\newcommand{\mbL}{\mathbf{L}}
\newcommand{\mbM}{\mathbf{M}}
\newcommand{\mbN}{\mathbf{N}}
\newcommand{\mbO}{\mathbf{O}}
\newcommand{\mbP}{\mathbf{P}}
\newcommand{\mbQ}{\mathbf{Q}}
\newcommand{\mbR}{\mathbf{R}}
\newcommand{\mbS}{\mathbf{S}}
\newcommand{\mbT}{\mathbf{T}}
\newcommand{\mbU}{\mathbf{U}}
\newcommand{\mbV}{\mathbf{V}}
\newcommand{\mbW}{\mathbf{W}}
\newcommand{\mbX}{\mathbf{X}}
\newcommand{\mbY}{\mathbf{Y}}
\newcommand{\mbZ}{\mathbf{Z}}

\renewcommand{\div}{\mrm{div}\,}
\newcommand{\divx}{\mrm{div}_{\mathbf{x}}\,}
\newcommand{\divy}{\mrm{div}_{\mathbf{y}}\,}
\newcommand{\nablax}{\nabla_{\mathbf{x}}}
\newcommand{\nablay}{\nabla_{\mathbf{y}}}
\newcommand{\rot}{\mrm{\mathbf{rot}}\,}
\newcommand{\curl}{\mrm{\mathbf{curl}}\,}
\newcommand{\curlx}{\mrm{\mathbf{curl}}_{\mathbf{x}}\,}
\newcommand{\curly}{\mrm{\mathbf{curl}}_{\mathbf{y}}\,}
\newcommand{\mL}{\mrm{L}}
\newcommand{\mH}{\mrm{H}}
\newcommand{\mV}{\mrm{V}}
\newcommand{\mW}{\mrm{W}}
\newcommand{\loc}{\mrm{loc}}
\newcommand{\out}{\mrm{out}}
\newcommand{\ext}{\mrm{ext}}
\newcommand{\supp}{\mrm{supp}}
\newcommand{\Ker}{\mrm{Ker}\,}
\newcommand{\coker}{\mrm{coker}\,}


\newcommand{\ftilde}{\ensuremath{\tilde{f}}}
\newcommand{\Uinc}{\ensuremath{U_{\text{inc}}}}
\newcommand{\Usc}{\ensuremath{U_{\text{sc}}}}
\newcommand{\IMean}{\ensuremath{I}}
\newcommand{\CMean}{\ensuremath{C}}

\newcommand{\ui}{\ensuremath{u_\mathrm{inc}}}
\newcommand{\ut}{\ensuremath{u_\mathrm{tot}}}
\newcommand{\us}{\ensuremath{u_\mathrm{sc}}}

\newcommand{\density}{\rho}

\renewcommand{\vec}[1]{\boldsymbol{#1}}
\newcommand{\unitvec}[1]{\hat{\vec{#1}}}

\newcommand{\eu}{\mathrm{e}}
\newcommand{\iu}{\mathrm{i}}

\newcommand{\mbrtilde}{\mathbf{\tilde{r}}}
\newcommand{\rtilde}{\tilde{r}}
\newcommand{\thetatilde}{\tilde{\theta}}
\newcommand{\vftilde}{\tilde{\vf}}

\newcommand{\ensem}[1]{\langle #1 \rangle}
\newcommand{\pc}{\mrm g}
\newcommand{\proba}{\mrm p}
\newcommand{\dd}{\mrm d}

\newcommand{\smallc}{\ensuremath{\mathfrak{c}}}
\newcommand{\BIGC}{\ensuremath{\mathfrak{C}}}


\numberwithin{equation}{section}

~\vspace{-3cm}
\begin{center}
{\sc \bf\LARGE Effective T-matrix of a cylinder  filled \\[6pt]  with a random 2D particulate}
\end{center}

\begin{center}
\textsc{K. K. Napal}$^1$, \textsc{P. S. Piva}$^{1}$ and \textsc{A. L. Gower}$^1$
\\[16pt]
\begin{minipage}{0.92\textwidth}
{\small
$^1$ Department of Mechanical Engineering, University of Sheffield, UK\\[10pt]
E-mails: \textit{k.k.napal@sheffield.ac.uk},\, \textit{pspiva1@sheffield.ac.uk}, \textit{a.l.gower@sheffield.ac.uk}\\[-14pt]
\begin{center}
(\today)
\end{center}
}
\end{minipage}
\end{center}
\vspace{0.4cm}

\thispagestyle{plain}

\noindent\textbf{Abstract.} When a wave, such as sound or light, scatters within a densely packed particulate, it can be rescattered many times between the particles, which is called multiple scattering. Multiple scattering can be unavoidable when: trying to use sound waves to measure a dense particulate, such as a composite with reinforcing fibers. Here we solve from first principles multiple scattering of scalar waves, including acoustic, for any frequency from a set of 2D particles confined in a circular area. This case has not been solved yet, and its solution is important to perform numerical validation, as particles within a cylinder require only a finite number of particles to perform direct numerical simulations. The method we use involves ensemble averaging over particle configurations, which leads us to deduce an effective T-matrix for the whole cylinder, which can be used to easily describe the scattering from any incident wave.
In the specific case when the particles are monopole scatters, the expression of this effective T-matrix simplifies and reduces to the T-matrix of a homogeneous cylinder with an effective wavenumber $k_\star$. To validate our theoretical predictions we develop an efficient Monte Carlo method and conclude that our theoretical predictions are highly accurate for a broad range of frequencies.
\\
\newline
\noindent\textbf{Subject Areas.} wave motion, acoustics, statistical physics
\\

\section{Introduction}

\noindent \textbf{Ensemble averaging.} Multiple scattering is unavoidable when using waves to characterise a particulate composite, or designing metamaterials to control wave propagation. Further, the number of particles in most applications makes direct numerical simulations impossible for current computing power, though there are some notable attempts \cite{gumerov2005computation,koc1998calculation}. Even if such simulations were possible, the positions of the particles are impossible to know for one particular sample. A mathematical method to solve this problem, called ensemble average, is to take an average over all possible particle positions. This type of averaging can occur naturally for fluids and gases, as the particles are moving. That is, taking an average of the scattered signal over time can be equivalent to an ensemble average\footnote{In ergodic systems, if enough time has passed, all physically possible states of the system will have occurred, and so that taking an average over time is equivalent to averaging over all possible configurations.}. Assuming that ensemble averaging is equivalent to averaging over time and space, which is often the case, is called the ergodicity assumption \cite{mishchenko2006multiple}.      


\vspace{10pt}

\noindent \textbf{What is known.} The scenarios that are best understood are: 1) waves in an infinite media with no boundaries \cite{sheng2007introduction,vynck2021light}, and 2) plane waves incidence on a half-space or plate filled with particles \cite{twersky1962scatteringI,twersky1962scatteringII}. Both scenarios have been considered to obtain effective wavenumbers \cite{foldy1945multiple,linton2005multiple,waterman1961multiple}. Though we mention the methods that use Lipmann-Schwinger for acoustics, they involve an extra integral which is often omitted \cite{martin2003acoustic} and complicates the calculations. Our approach in this work is valid for any type of scatterer and scalar waves.
Both scenarios of using plane waves and an infinite media have several applications (typically when considering layered media such as planetary atmospheres, layers in the ocean, or soils) but one significant drawback is that it has been very challenging to numerically validate the assumptions used for these methods. Both use statistical assumptions, such as the Quasi-Crystalline Assumption (QCA) \cite{lax1952multiple} that is not based on an asymptotic approximation. Validation is needed to establish the range of validity of these assumptions. However, direct numerical simulations of scattering from a configuration of particles for both planes and infinite media require a huge number of particles \cite{karnezis2023average},  \cite{gower2018characterising} or the introduction of periodic boundaries which can introduce artifacts \cite{chekroun2012time}.
\vspace{10pt}

\noindent \textbf{The cylindrical setting.} The methods developed to describe the average plane wave propagating in a disordered particulate plate or half-space can now be extended to other geometries \cite{gower2021effective}. The ideal scenario to compare theoretical predictions with direct numerical simulations is to have cylindrical particles inside a cylinder, as this reduces the problem to two dimensions and we need only a finite number of particles for the direct numerical simulations. See Figure~\ref{fig:modal scattering} for an illustration. This is the simplest case to perform numerical validation of a very general theory \cite{gower2021effective}. Further, we show in this work that the effective dispersion equation for the cylindrical geometry is the same as the plane-wave case. So numerically validating the cylindrical geometry will also serve as numerical validation of the dispersion equation for plane waves and all geometries. We also note that it appears that the cylindrical setting has never been solved from a first principles approach. The formulas we provide are also valid for any inter-particle pair correlation.

\vspace{10pt}

\noindent \textbf{Industrial applications.} Beyond numerical validation, there are industrial applications that need a method to calculate waves scattered from a cylinder with particles. Examples of cylinders filled with cylindrical particles include concrete beams reinforced with iron, cables filled with wires, or fibre-reinforced composite \cite{bose1973longitudinal}. Applications include designing cylinders with exotic effective properties or developing methods to measure the cylindrical particles \cite{ni2010achieving,pendry2004reversing}. In terms of measurement, it is likely that more information can be extracted from waves scattered from a cylinder filled with a particulate than just plane wave reflection from a plate filled with the same particulate.  
\vspace{10pt}


\noindent \textbf{Effective properties.} The most common approach to model the average scattering from, say, a spherical or cylindrical region with particles is to assume the region is homogeneous with some effective properties \cite{torrent2006homogenization,rohfritsch2019numerical,Dubois2011,linton_multiple_2005,Tishkovets2011,Tsang1982,Varadan1983}, and then use the standard boundary conditions such as continuity of displacement. This approach is valid for low-frequency \cite{torrent2006homogenization} but for finite frequencies is incorrect in three dimensions \cite{gower2021effective}, and we demonstrate the same here for two dimensions in this work. To obtain an accurate model for a finite frequency the boundary condition needs to be deduced from first principles, together with an eigensystem for the effective wavenumbers \cite{gower2021effective}. Although the process is more involved, the final expression for the average scattered wave from a cylindrical region is simple: the average scattered wave can be calculated from an effective T-matrix for any incident frequency, source, and particle properties. We stress that without deducing the results from first principles, as we do here, it would not be possible to just guess the form of this effective T-matrix.
\vspace{10pt}

\noindent \textbf{Monopole scatterers.} One somewhat surprising result we deduce is that if the particles scatter only monopole waves, that is waves that have radial symmetry, then the material as a whole behaves as homogeneous, where the mass density is the same as the background, and the bulk density is given by a simple formula. We deduce this for particles in a cylinder and hypothesize that it is true for any material filled with monopole scatterers, even when including all orders of multiple scattering. Beyond just curiosity, there are many particles that behave approximately like a monopole scatterer, and therefore the simple formulas we deduced are appropriate. For example, in acoustics void-like particles are approximately monopoles for a broad frequency range, see the Dirichlet case in Figure \ref{fig:intro MC result}. In elasticity, particles become approximately monopole when the bulk modulus is much greater than the shear modulus  \cite{cotterill2022deeply}. Other cases include resonators such as a split ring resonator \cite{smith2022tailored}.
\vspace{10pt}

\noindent \textbf{Overview of the method.} 
After ensemble averaging over particle configurations within a cylindrical region, the system inherits cylindrical symmetry. For example, if the source has radial symmetry then the average scattered field will also have radial symmetry. This is also true for sources with more general rotational symmetry resulting in scattered fields of the same rotational symmetry (cf. Figure~\ref{fig:modal scattering}). In this paper, we take advantage of this mode-to-mode symmetry to analyze the general behaviour of the random particulate material independently from the incident field. 

After denoting with $\mrm V_n$ and $\mrm U_n$ the regular and outgoing cylindrical waves of order $n$ (cf. \ref{def:u_n and v_n for a cylinder}), the cylindrical symmetry translates as follows: when the exciting source is $\mrm V_n$, then the average scattered field is $\mrm T_n \mrm U_n$ where the complex number $\mrm T_n$ only depends on the properties of the random particulate cylinder (radius and properties of the particles). Since the scattering problem is linear, the knowledge of the $\mrm T_n$ allows us to describe the scattering from any incident field, after decomposing the latter into the modes $\mrm V_n$. Having simple expressions for $\mrm T_n$ is crucial to help guide methods to characterize or design particulate materials. We do so by using the effective wave method approach \cite{gower2021effective}. Finally, we validate our results with an adapted Monte Carlo method in which the rate of convergence is accelerated thanks to the cylindrical symmetry. 
\vspace{10pt}

\begin{figure}[ht]
\fcolorbox{gray!20}{gray!20}{
\begin{minipage}{.999\linewidth}
    \centering
    \textbf{The mode to mode scattering}\\[5pt]
\includegraphics[scale=.32,trim={1cm 0cm 0cm 0cm},clip]{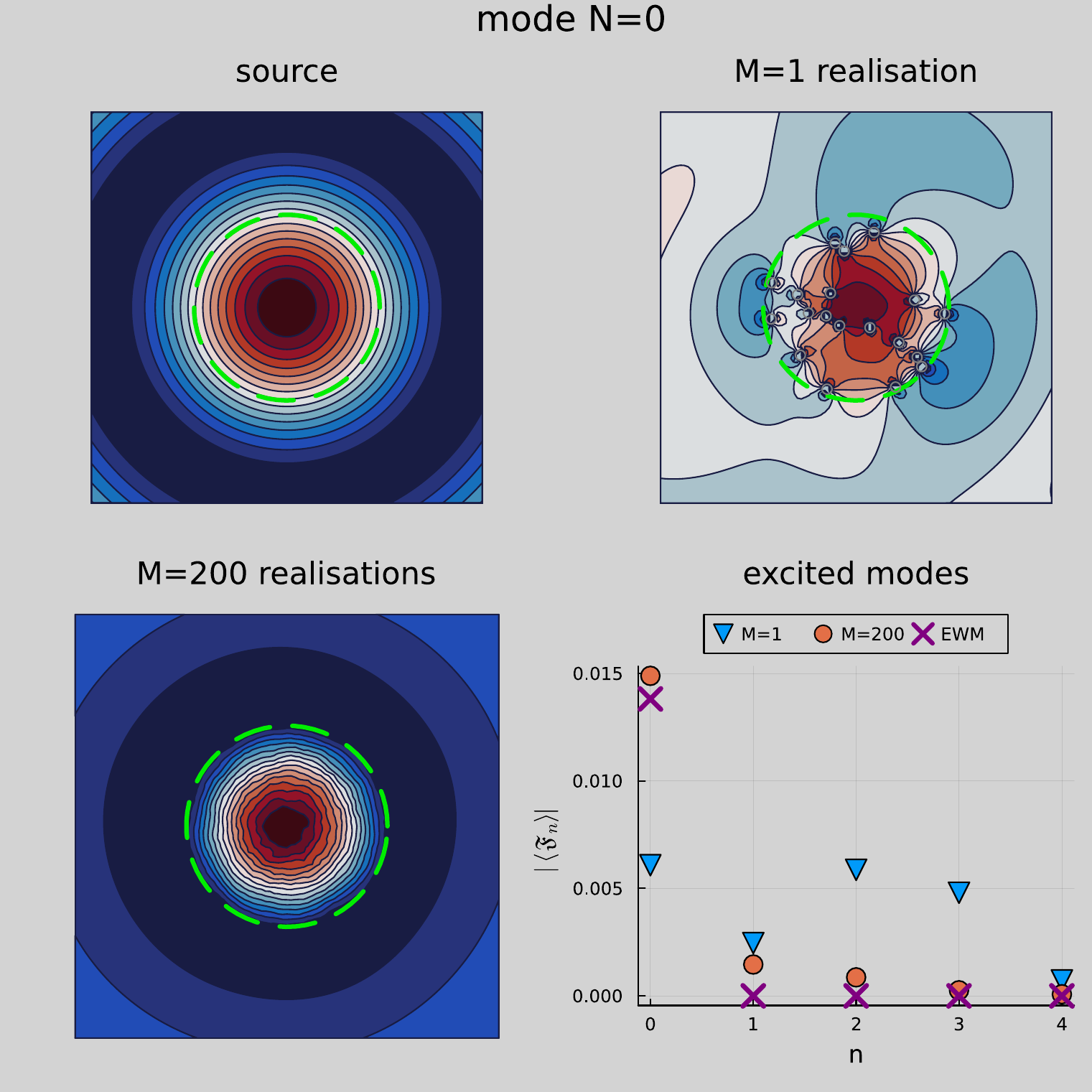}
\hfill
\includegraphics[scale=.32,trim={1cm 0cm 0cm 0cm},clip]{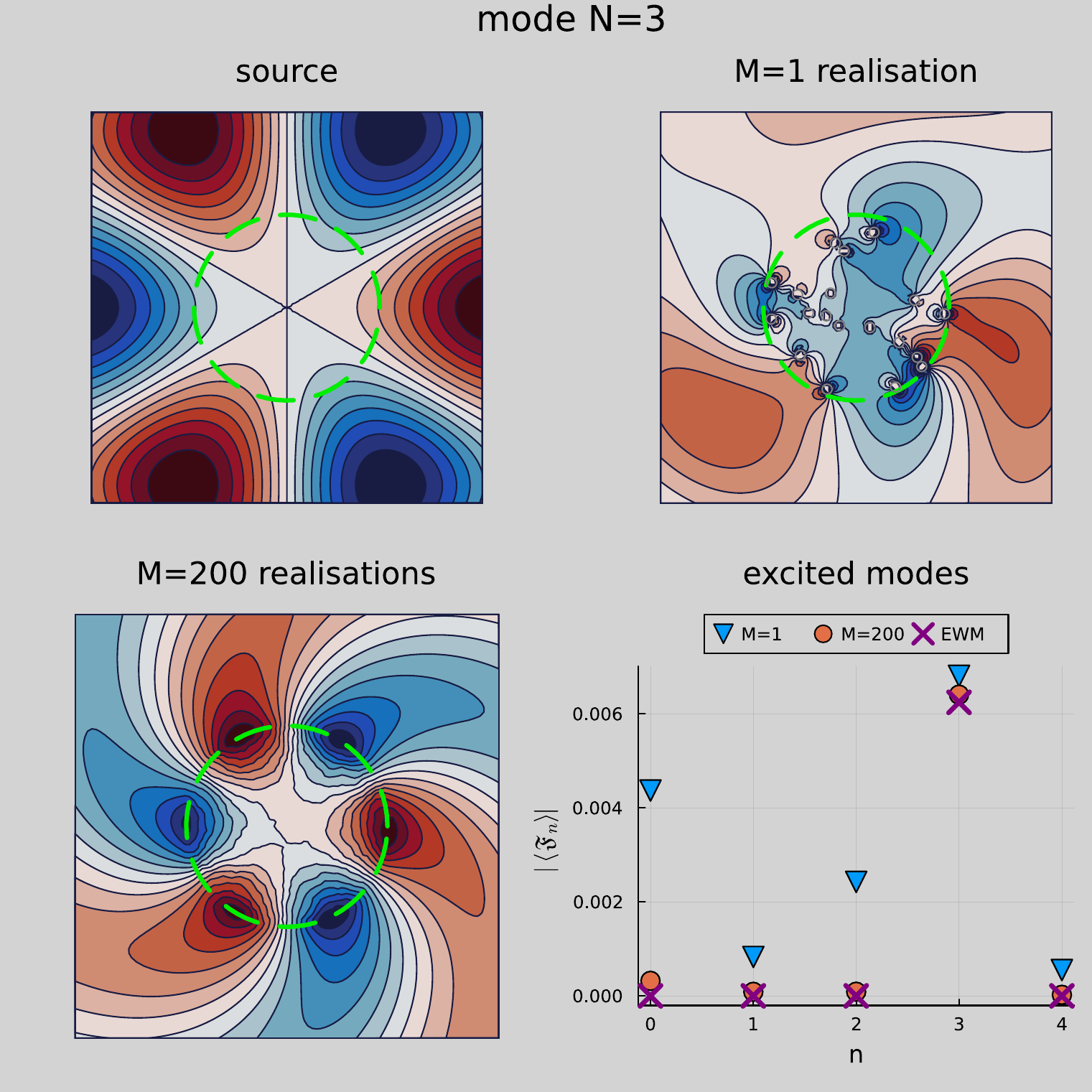}
       \caption{ Consider a circular region filled with sound-hard particles. Both panels compare the average scattered wave $\ensem{\us}$, for a source with only one mode: $\ui=\mrm V_N(\mbr)$, when using our effective wave method and a brute force Monte Carlo approach. The Monte Carlo approach simulates the scattered field from either one, $M=1$, or two hundred, $M=200$, configurations of particles and then takes the average over these configurations. In general, the average scattered wave is given by \eqref{eq:averaged us}.  
       The left panel shows that when using an incident wave $\ui=\mrm V_0(\mbr)$, with radial symmetry, then the only scattered mode also has radial symmetry after average. The right panel shows how a source with a $60^\circ$ rotation symmetry also leads to a scattered wave with the same symmetry.        
       }
    \label{fig:modal scattering}
\end{minipage}\hfill
}
\end{figure}


\noindent \textbf{Overview of this paper}  

In Section \ref{sec:Definitions}, we first introduce the statistics of the random particulate material and the required notations for the ensemble averaging. We then define the T-matrix of the effective cylinder whose exact formula depends on the solutions of the averaged Foldy-Lax equations.

The latter is solved in Section  \ref{sec:effective waves method} by using the effective waves method, which consists in finding solutions that are isotropic waves with a complex wavenumber $k_\star$. The method leads to an eigenvalue problem called the dispersion equation, whose eigenvalue provides $k_\star$.

In Section \ref{sec:effective T-matrix}, we use the expression of the solutions of the averaged Foldy-Lax equations to deduce a formula of the effective T-matrix. The latter is very simple when the particles are monopole scatterers:
\begin{equation}
\label{eq:intro cylinder T-matrix (monopole)}
     \mrm T_n = 
    - \frac{
  \Cc_n
  }
  {
 \Dc_n
  }
  \qquad
  \text{with}
  \qquad
  \left\{
    \begin{array}{rcl}
  \Cc_n &=&\dsp
k\mrm J'_{n}(k\tilde{R})\mrm J_{n}(k_\star\tilde{R})-k_\star\mrm J_{n}(k\tilde{R})\mrm J'_{n}(k_\star\tilde{R})\\
  \Dc_n&=& \dsp
k\mrm H'_{n}(k\tilde{R})\mrm J_{n}(k_\star\tilde{R})-k_\star\mrm H_{n}(k\tilde{R})\mrm J'_{n}(k_\star\tilde{R}) 
  \end{array}
    \right.
\end{equation}
where $\tilde{R}:=R-a$. This result is remarkable because the above expression corresponds to the T-matrix of a homogeneous acoustic cylinder of radius $\tilde{R}$, sound speed $c_\star=\omega/k_\star$ and density $\rho$ which is equal to the background medium.  Particles that are approximately monopole scatterers appear in mainly two scenarios: either small sound soft particles or resonators \cite{smith2022asymptotics}. For all these cases, the T-matrix~\eqref{eq:intro cylinder T-matrix (monopole)} shows us that the effective wavenumber $k_\star$ suffices to describe the random material.

When the particles are not monopole scatterers, \eqref{eq:intro cylinder T-matrix (monopole)} is not exact. The exact formula is given by
\begin{equation}
\label{eq:intro cylinder T-matrix}
\mrm T_{n} =- \frac{
\dsp \sum_{n'}  \Cc_{n-n'}F_{n'}
}
{
\dsp\sum_{n'} \Dc_{n-n'}F_{n'}}
\end{equation}
where the $\Cc_n$ and $\Dc_n$ are the same as before, and the weights $F_n$ are the eigenfunctions of the dispersion equation, associated with the effective wavenumber $k_\star$.

For monopole scatterers, we have that ($n'=0$), and the above reduces to \eqref{eq:intro cylinder T-matrix (monopole)}. We note that often $F_{0}$ is the largest term, which explains why \eqref{eq:intro cylinder T-matrix (monopole)} can give accurate results for non-monopole scatterers, see for example the numerical results for sound soft (Dirichlet) particles shown in Figure~\ref{fig:intro MC result} in the low-frequency regime. In this same figure, we also see how for sound hard particles (Neumann), which are dipole scatters, the Monte Carlo results and the formula~\eqref{eq:intro cylinder T-matrix} closely match, whereas the T-matrix for monopole scatters doesn't match.
\vspace{10pt}

\begin{figure}[h]
\fcolorbox{gray!20}{gray!20}{
\begin{minipage}{.999\linewidth}
    \centering
\textbf{Validation of \ref{eq:intro cylinder T-matrix (monopole)} and \ref{eq:intro cylinder T-matrix}}\\[5pt]
    \includegraphics[scale=.63,trim={0 0 0 1.87cm},clip]{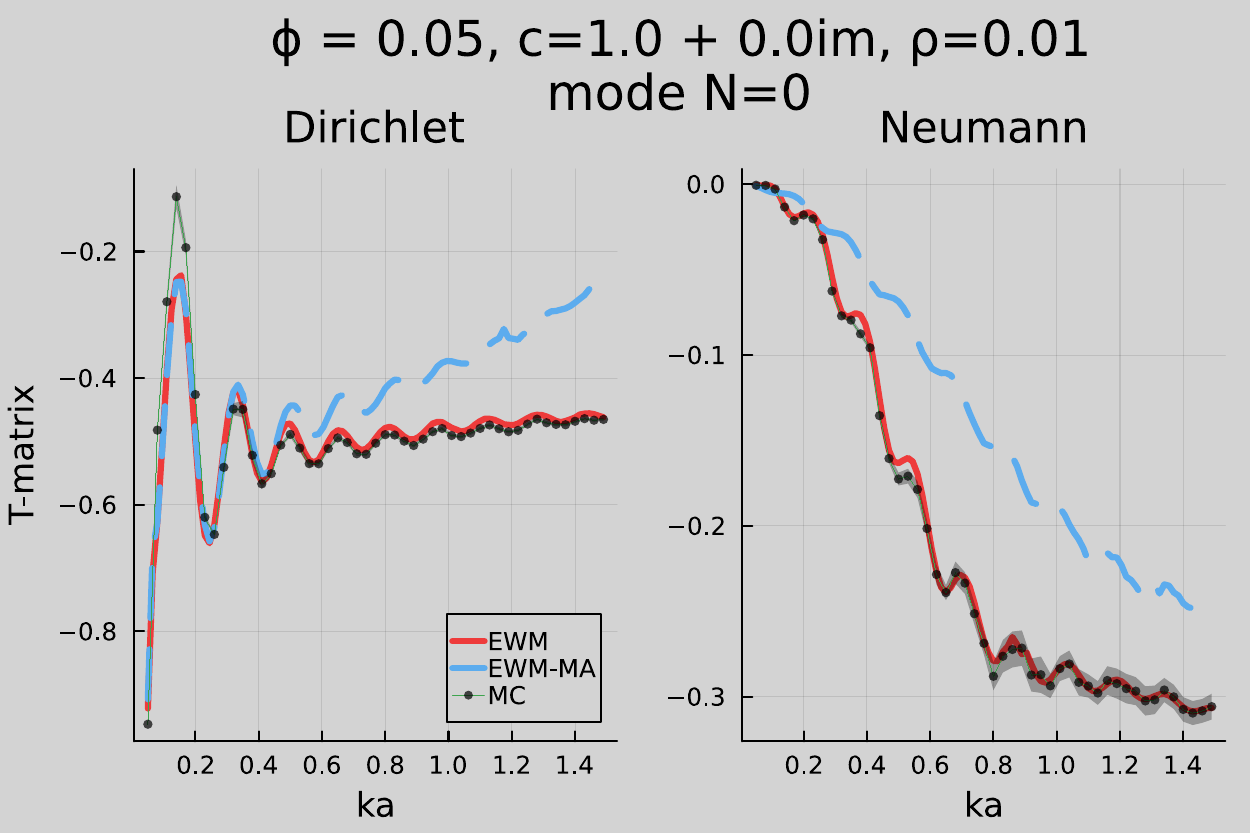}
    \caption{Compares various methods to calculate the component $\mrm T_0$ of the T-matrix of a cylinder filled with particles. The solid red line is our effective waves method (EWM) \eqref{eq:effective T-matrix}, the black points are from a brute force Monte Carlo method (MC), and the dashed blue line is our method when only monopole scattering is accounted for (EWM-MA) \eqref{eq:intro cylinder T-matrix (monopole)}. The left (right) graph shows the results for sound soft (hard) particles. In both cases, the general expression of the effective T-matrix matches the Monte Carlo results. The EWM-MA method only matches well with the Monte Carlo for sound soft scatterers for low frequencies.}
    \label{fig:intro MC result}
  \end{minipage}\hfill
  }
\end{figure}

\begin{figure}[ht]
\fcolorbox{gray!20}{gray!20}{
\begin{minipage}{.999\linewidth}
    \centering
\textbf{Set up of cylindrical particles inside a cylinder}\\[5pt]
\TikzCylinders{.65}
\caption{Illustration of a possible configuration of four particles. The particles are cylinders whose axes are aligned with the $z$-axis. The position $\mbr_i\in\R$ of the $i^\text{th}$ particle is determined by the intersection of its axis with the plane $xOy$. The particles are of radius $a$ and physically contained in a cylinder of radius $R$. In this specific example, all the particles are the same and the centres $\mbr_i$ are therefore confined in the same disc $\Rc_1$  shown in green.}
\label{fig:settings}
\end{minipage}\hfill
}
\end{figure}

\section{Random particulate material}
\label{sec:Definitions}

\subsection{Deterministic scattering from J particles}
Here we summarise some results for multiple scattering of acoustic waves, in the time-harmonic regime, by a collection of $J$ cylinders, referred as to particles, which axis are aligned with the $z$-axis. The centre of the particle $i$ is identified by $\mbr_i\in\R^2$ as shown in Figure \ref{fig:settings} and is assumed to be confined in an area denoted by $\Rc_i$.

The propagation of waves in free space is governed by the 2D Helmholtz equation 
\begin{equation}
\label{eq:Helmholtz}
    \Delta u + k^2u= 0 
\end{equation}
where $\Delta:=\d_x^2+\d_y^2$ is the 2D Laplace operator and  $k\in\R$ is the wavenumber of the homogeneous background. Consider the following two independent bases of the Helmholtz equation $\mrm V_n$ and $\mrm U_n$ defined by
\begin{equation}
\label{def:u_n and v_n for a cylinder}
\begin{array}{|rcll}
\mathrm U_n(k \mbr)&:=& \mathrm H_n(k r)\mathrm e^{\mrm in\theta} & \forall \mbr\in \R^2\setminus\{\bm{0}\} 
\\
\mathrm V_n(k\mbr)&:=&\mathrm J_n(kr)\mathrm e^{\mrm in\theta} & \forall \mbr\in\R^2
\end{array}
\end{equation}
where $k$ is the background wavenumber and  $(r,\theta)$ are the polar coordinates of $\mbr$, ie $\mbr = (r\cos \theta,r\sin\theta)$. The specific solutions $\mrm V_n(k\mbr)$ have the particularity of being smooth while $\mrm U_n(k\mbr)$ have a singularity at the origin and are outgoing solutions. 

Both the incident field $\ui$ and the scattered field $\us$ are solutions of \eqref{eq:Helmholtz}. While we assume that  $\ui$ is smooth and regular in the region that covers the particles\footnote{This is true for any source which originated from outside this region.} the scattered field has to be a sum of outgoing fields from each particles centered at $\mbr_i$, as a result we can write
\vspace{-20pt}
\begin{subequations}\label{eq:inc and sc fields intro}
    \begin{multicols}{2}
        \begin{equation}
          \label{eq:incident field intro}
 \ui(\mbr) =  \sum_{n=-\infty}^{+\infty}g_n\mrm V_n(k\mbr)
        \end{equation}\break
        \begin{equation}
            \label{eq:scattered field intro}
             \us(\mbr) =  \sum_{i=1}^J\sum_{n=-\infty}^{+\infty} f_n^i\mrm U_n(k\mbr-k\mbr_i).
        \end{equation}
    \end{multicols}
\end{subequations}
\noindent
The coefficients $g_n\in\C$ are given parameters of the system and $f_n^i\in\C$ are unknowns that can be determined by following the Foldy-Lax self-consistent method \cite{martin2006multiple}. The latter leads to the following system of equations 
\begin{equation}
\label{eq:f_n^i governing equation}
f_n^i=\mrm T_n^i\sum_{n'}\mrm V_{n'-n}(k\mbr_i)g_{n'}
+ \mrm T_n^i\sum_{j\neq i}  \sum_{n'} \mrm U_{n'-n}(k\mbr_i-\mbr_j)f_{n'}^j  \quad \forall n\in\Z,\,\forall i=1\dots J,
\end{equation}
where $\mrm T_n^i$ is the T-matrix of particle $i$, which can represent a wide range of particles \cite{ganesh2017algorithm}. 

To give an example, the expression of a T-matrix of a homogeneous particle with wavenumber $k_i$, density $\rho_i$ and radius $a_i$ is given by
\begin{equation}
 \label{eq:T-matrix-acoustics}
\mrm T^i_{n} = 
-\frac{\density_ik \mathrm J_n' (k a_i) \mathrm J_n (k_i a_i)-\density k_i\mathrm J_n (k a_i) \mathrm J_n' (k_i a_i)}
{\density_ik\mathrm H'_n (k a_i) \mathrm J_n (k_i a_i)- \density k_i\mathrm H_n (k a_i) \mathrm J_n' (k_i a_i)}\cdot
\end{equation}

The system \eqref{eq:f_n^i governing equation} totally determines the scattering coefficients $f_n^i$, which allows to solve the scattering problem from the $J$ given particles. However, this result is not very useful in practice for two main reasons: first, the position of the particles is often unknown, and second, there can be a very large number of particles in most industrial applications \cite{challis2005ultrasound}. The ensemble average over all particle positions, that we summarize below, solves both these problems from the computational standpoint.


\subsection{Particulate distribution}


We describe each particle, say particle $i$, with two random variables: $\mbr_i$ is the probability distribution of the centre of the particle in space, and $\lambda_i\in\Sc$ describes all other properties of the particle (radius $a_i$, density $\rho_i$, etc.). 


The centre of a particle $\mbr_i$ is restricted in a region $\Rc_i$ which depends on the particle's properties $\lambda_i$. For example, $\mbr_i$ has to be one radius away from the boundary $\d\Rc$ which completely contains all particles. In this paper we assume that $\mbr_i$ is distributed uniformly over the set $\Rc_i(\lambda_i)$ defined by
\begin{equation}
    \Rc_i:=\{\mbr\in\Rc\,:\,\mrm{dist}(\mbr,\d\Rc)>a_i\}.
\end{equation}
As a consequence,
\begin{equation}
\label{eq:particle position uniform distribution}
    \proba(\mbr_i,\lambda_i) =  \proba(\mbr_i|\lambda_i) \proba(\lambda_i) =\frac{\proba(\lambda_i)}{|\Rc_i|}, 
\end{equation}
where $|\Rc_i|$ is the volume of $\Rc_i$.

\subsubsection*{Joint particle distribution}

Each particle is described by two random variables $(\lambda_i,\mbr_i)$, $i=1,\dots, J$.
In general, the particle positions are correlated, for example, no particles can overlap. The most commonly used term to describe inter-particle correlation is the pair correlation $\pc$, which satisfies (cf. \cite[eq. 8.1.2]{kong2004scattering}):
\begin{equation}
\label{def:pair correlation}
 \proba(\mbr_1;\mbr_2|\lambda_1;\lambda_2) = \frac{\pc(\mbr_1, \lambda_1; \mbr_2,  \lambda_2)}{|\Rc_1||\Rc_2|} \frac{ J}{ J-1},
\end{equation}
 where on the left is the joint law of two particles' position when their properties are known. 
 
 The pair correlation $\pc$ describes how correlated any two particles are, when the positions and properties of all other particles are unknown. For example, if $\pc = 1$ for all values of its arguments then both $\mbr_i$ and $\mbr_j$ in the above are independent and uniformly distributed over $\Rc_i$ and $\Rc_j$, respectively (in the limit $J\to\infty$). 
 
Finally, we introduce the density  
\begin{equation}
\label{def:mathfrak n}
    \mathfrak{n}(\lambda_i):=\frac{ J}{|\Rc_i|}\proba(\lambda_i)
\quad    
    \text{(number of $\lambda_i$ types particles per unit volume).}
\end{equation}
Then we derive the following useful relation
%
%
\begin{equation}
\label{eq:conditional probability}
    \proba(\mbr_j,\lambda_j|\mbr_i,\lambda_i) 
    = \frac{ \proba(\mbr_j,\lambda_j;\mbr_i,\lambda_i) }{ \proba(\mbr_i,\lambda_i) } 
    = |\Rc_i|\proba(\lambda_j)\proba(\mbr_i;\mbr_j|\lambda_i;\lambda_j)
    = \frac{\mathfrak{n}(\lambda_j)}{ J - 1}\pc(\mbr_i,\mbr_j),
\end{equation}
where we used \eqref{eq:particle position uniform distribution}, \eqref{def:pair correlation} and \eqref{def:mathfrak n}. 

\subsubsection*{The pair correlation}

In this paper, we consider that the particles have a distribution which is isotropic and homogeneous in space. As a consequence, the pair correlation is of the form $\pc(\mbr_i, \lambda_i;\mbr_j, \lambda_j) = \pc(|\mbr_j-\mbr_i|, \lambda_i, \lambda_j)$. We assume the pair correlation is of the following form: 
\begin{equation} \label{eqn:split-pc}
\pc(r,\lambda_1,\lambda_2) =
    \begin{cases}
    0, \quad &r < a_{12}, \\ 
    1 + \delta \pc(r,\lambda_1,\lambda_2), \quad &  a_{12} < r  < b_{12}, \\
    1, \quad &  r  > b_{12},
    \end{cases}
\end{equation}
where $g(r,\lambda_1,\lambda_2) = 0$ when particles overlap, with $a_{12} \geq a_1 + a_2$ and in general $a_{12}$ can depend on $\lambda_1$ and $\lambda_2$. This form of the pair correlation moreover assumes that at a certain distance $b_{12}$ from each other, the particles become uncorrelated, that is $g(r,\lambda_1,\lambda_2) = 1$ for $r>b_{12}$. This assumption will lead to analytic simplifications, as well as being a good approximation for most disordered materials. A typical plot of the pair correlation is illustrated in Figure \ref{fig:pc illustration}.

%

\begin{figure}[h]
\centering
\fcolorbox{gray!20}{gray!20}{
\begin{minipage}{.65\linewidth}
\begin{center}
\textbf{Radial pair-correlation function}\\[5pt]
\includegraphics[scale=.45]{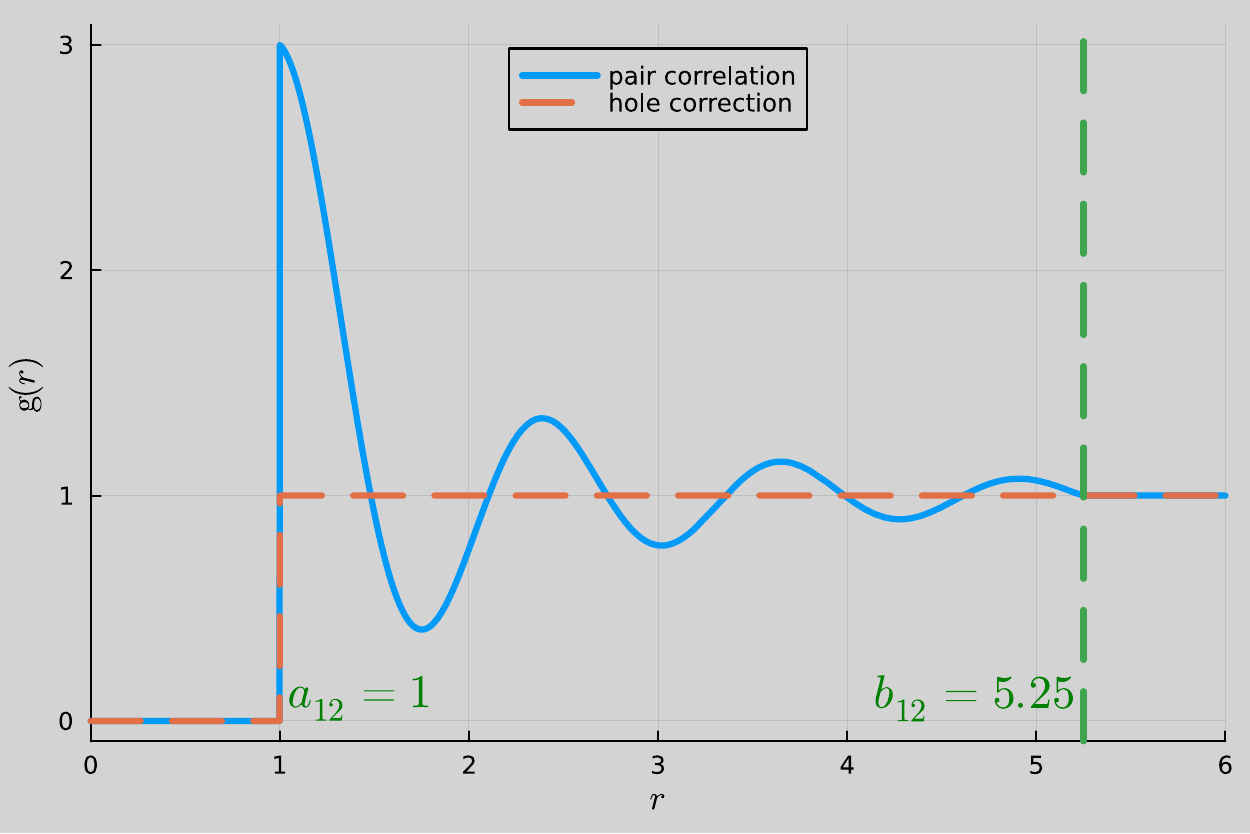}
\caption{Typical plot of the pair correlation function. }
\label{fig:pc illustration}
\end{center}
\end{minipage}
\begin{minipage}{.34\linewidth}
The pair-correlation $\pc(r)$ is zero for $r < a_{12}$ because particles cannot overlap. The local minima (maxima) indicate that there is a distance $r$ from one fixed particle, say at $\mbr_1$, where it is less (more) likely to find other particles. Finally, assuming $\pc(r)=1$ for $r \geq b_{12}$ means that a particle at $\mbr_1$ becomes uncorrelated to particles which are further than $b_{12}$.
\vspace{30pt}
\end{minipage}
}
\end{figure}

\subsection{Definition of the effective T-matrix}

The scattered field \eqref{eq:scattered field intro} can be simplified after using Graf's addition theorem (\ref{eq:translation identities},ii) with $\mbx=k\mbr$ and $\mbd=-k\mbr_i$
\vspace{-20pt}
\begin{subequations}\label{eq:us multimodal decomposition}
    \begin{multicols}{2}
        \begin{equation}
          \label{eq:us compact expression}
 \us(\mbr) =  \sum_{n=-\infty}^{+\infty} \mathfrak{F}_n\mrm U_n(k\mbr)
        \end{equation}\break
        \begin{equation}
            \label{def:mathfrak Fn}
 \mathfrak{F}_n :=  \sum_{i=1}^J\sum_{n'=-\infty}^{+\infty}\mrm V_{n'-n}(-k\mbr_i)f_{n'}^i.
        \end{equation}
    \end{multicols}
\end{subequations}
%
\noindent
Note that \eqref{eq:us multimodal decomposition} is only valid for $r>R$, nevertheless this is enough for the definition of the effective T-matrix below.
Taking the ensemble average of the equation above, as defined in Apendix \ref{app:ensemble_average}, leads to  
%
\begin{equation}
\label{eq:averaged us}
\ensem{\us(\mbr)} =  \sum_{n=-\infty}^{+\infty} \ensem{\mathfrak{F}_n}\mrm U_n(k\mbr).
\end{equation} 
Since the scattering problem is linear with respect to the incident field, $\ensem{\mathfrak{F}_n}$ depends linearly on the modes $g_n$ of the incident field \eqref{eq:incident field intro}:
\begin{equation}
\label{def:effective T-matrix}
 \tcboxmath{
\ensem{\mathfrak{F}_n} = \sum_{N=-\infty}^{+\infty}\Tc_{n,N} g_N \quad \text{(effective T-matrix definition)}.
}
\end{equation}
This relation defines the T-matrix $\Tc$ of the averaged material which connects the modes of the incident field with the ones of the averaged scattered field. It allows us to describe the scattering from any incident field, provided the coefficients $g_n$ in \eqref{eq:incident field intro}.

In the specific case when the area where the particles are confined is circular, we have $\mathcal T_{n,N}=0$ if $n\neq N$ (cf. Appendix \ref{app:T is diagonal}), so that \eqref{eq:averaged us} becomes 
\begin{equation}
\label{eq:averaged scattered field (circular region)}
\ensem{\us(\mbr)} =  \sum_{n=-\infty}^{+\infty} \mathrm T_ng_n\mrm U_n(k\mbr).
\end{equation}
where $\mathrm T_n:=\mathcal T_{n,n}$.
\vspace{10pt}

Computing $\Tc_{n,N}$ requires to compute $\ensem{\mathfrak{F}_n}$. From \eqref{eq:us compact expression}, \eqref{def:ensemble average}, and \eqref{eqn:compute <Fn>} we obtain  %
\begin{equation}
\label{eq:averaged scattering coefficients}
\ensem{\mathfrak{F}_n} = \int_\Sc\mathfrak{n}(\lambda_1)\int_{\Rc_1}\sum_{n'} \mrm V_{n'-n}(-
k\mbr_1)\ensem{f_{n'}}(\mbr_1,\lambda_1)\,\dd\mbr_1\dd\lambda_1.
\end{equation}
Here we used \eqref{eqns:indistinguishable} to substitute $\ensem{f_{n}}(\mbr_1,\lambda_1)=\ensem{f_{n}^1}(\mbr_1,\lambda_1)$. 

The function $\ensem{f_{n}}(\mbr_1,\lambda_1)$ needs to be determined before we can compute $\ensem{\mathfrak{F}_n}$. In Appendix~\ref{app:average governing equation computation} we show how to obtain the governing equation:
\begin{multline}
\label{eq:average governing equation}
\ensem{f_n}(\mbr_1,\lambda_1)
=   \mrm T_n(\lambda_1)\dsp \sum_{n'}\mrm V_{n'-n}(k\mbr_1)g_{n'} 
\\
+\mrm T_n(\lambda_1)\dsp\sum_{n'} \int_\Sc \mathfrak{n}(\lambda_2) \int_{\Rc_2} \mrm U_{n'-n}(k\mbr_1 - k \mbr_2)\ensem{f_{n'}}(\mbr_1,\lambda_1) \pc(|\mbr_1 - \mbr_2|, \lambda_1\lambda_2) \,\dd \mbr_2\dd\lambda_2,
\end{multline}
where we also used the simpler pair-correlation~\eqref{eqn:split-pc}.

In the next section, we use the \emph{effective wave method} to solve \eqref{eq:average governing equation}. This method introduced in \cite{gower2021effective} proved to be successful in 3D and provides a closed formula for $\ensem{f_n}(\mbr_1,\lambda_1)$. This in turn allows us to compute $\ensem{\mathfrak F_n}$ \eqref{eq:averaged scattering coefficients} and reach an explicit formula for $\mathcal T_{n,N}$ by specifying $g_n=\delta_{n,N}$ in \eqref{def:effective T-matrix}.

\section{The effective wave method}
\label{sec:effective waves method}

To solve the general integral governing equation \eqref{eq:average governing equation}, we use the effective waves method \cite{gower2021effective} as summarised below.

\vspace{0.1cm}
\noindent
\fcolorbox{black}[HTML]{E9F0E9}{\parbox{\textwidth}{%
\begin{center}
    \textbf{Overview of the effective waves method}
\end{center}
The starting point is to assume that there exists  $k_\star\in \C$ such that
\begin{equation}
\label{def:effective wavenumber assumption}
(\Delta + k_\star^2)\, \ensem{f_{n}}(\mbr_1,\lambda_1) = 0.
\end{equation}
This assumption will greatly simplify the governing integral equation \eqref{eq:average governing equation}, from which we will be able to determine both $k_\star$ and $ \ensem{f_{n,N}}(\mbr_1,\lambda_1)$. To summarise, the method has three steps:
\vspace{10pt}


\begin{enumerate}
\item \textbf{Separate microstructure and boundary.}  The assumption \eqref{def:effective wavenumber assumption} is used in the governing equation \eqref{eq:average governing equation} to derive two separate equations called the \textit{ensemble wave equation} and the \textit{ensemble boundary conditions}. The first one only depends on the microstructure of the random material while, in contrast, the second one takes into account the incident field and the shape of the random material, acting much like a boundary condition.
\vspace{5pt}
\item \textbf{The effective eigensystem.} We decompose $\ensem{f_{n,N}}(\mbr_1,\lambda_1)$ in the basis of functions $\mrm V_n(k_\star\mbr_1)$ and substitute the decomposition into the \textit{ensemble wave equation}. The coefficients $\mathbf{F}(\lambda_1)$ of the decomposition are then shown to be the eigenfunctions of an eigensystem which eigenvalue is $k_\star$.
\vspace{5pt}
\item  \textbf{The ensemble boundary condition.} To determine the amplitudes of the eigenfunctions $\mathbf{F}(\lambda_1)$ we then use the \textit{ensemble boundary condition}, which takes into account the incident field and the shape of the random material. 
\vspace{7pt}
\end{enumerate}}}

%

\subsection{Separate microstructure and boundary}

We follow the steps described above to determine the solutions $\ensem{ f_{n}}(\mbr_1,\lambda_1)$  of \eqref{eq:average governing equation}. The first step uses \eqref{def:effective wavenumber assumption}, and some algebraic manipulations shown in Appendix \ref{app:derivation of EWE and EBC} to rewrite the governing equation~\eqref{eq:average governing equation} into two separate equations:
\begin{align}
\label{eq:ensemble wave equation}
    &\tcboxmath{
\ensem{f_{n}}(\mbr_1,\lambda_1) +
    \sum_{ n'\in\Z}\mrm T_n(\lambda_1)\int_\Sc 
    \left[
    \frac{\Jc_{n'n}(\mbr_1)}{k^2-k_\star^2} 
    -\Kc_{n'n}(\mbr_1)
    \right]
    \mathfrak{n}(\lambda_2)\,\dd\lambda_2
    =0
    \;\; \text{(ensemble wave equation)}
    }
\\ \label{eq:ensemble boundary equation}
  &  \tcboxmath{
\sum_{n'} \mrm V_{n'-n}(k\mbr_1) g_{n'}+ \sum_{n'}\int_\Sc \frac{\Ic_{n'n}(\mbr_1)}{k^2-k_\star^2} \mathfrak{n}(\lambda_2)\,\dd\lambda_2= 0
    \qquad \text{(ensemble boundary condition).}
    }
\end{align}
The terms $\Jc_{n'n}(\mbr_1)$, $\Ic_{n'n}(\mbr_1)$, and $\Kc_{n'n}(\mbr_1)$, respectively defined in \eqref{def:matcal I and J} and \eqref{def:matcal K}, involve the function $\ensem{f_{n,N}}(\mbr_1,\lambda_1)$.

One of the key advantages of splitting the integral equation \eqref{eq:average governing equation} into the two separate equations above is that the ensemble wave equation \eqref{eq:ensemble wave equation} does not depend on the shape of the region $\Rc$ nor the incident wave. As we will see below, the ensemble wave equation \eqref{eq:ensemble wave equation} can be used to determine the effective wavenumber $k_\star$, which implies that $k_\star$ only depends on the microstructure: the density of the particles $\mathfrak{n}(\lambda_1)$, their properties provided by the T-matrix $\mrm T_n(\lambda_1)$, and the pair correlation $\pc$ which explicitly appears in the quantity $\Kc_{n,n'}(\mbr_1)$ \eqref{def:matcal K}. We further discuss how to interpret $k_\star$ in Section \ref{sec:kstar interpretation}.

On the other hand, the ensemble boundary condition \eqref{eq:ensemble boundary equation} acts like a boundary condition, and shows how the incident wave and material boundary affect the overall solution.

\subsection{The effective eigensystem}

Since  $\ensem{f_{n}}(\mbr_1,\lambda_1)$ satisfies \eqref{def:effective wavenumber assumption}, it can be decomposed into the modes
\begin{equation}
\label{eq:fn effective modes decomposition}
\ensem{f_{n}}(\mbr_1,\lambda_1) = \sum_{n_1}F_{nn_1}(\lambda_1)\mrm V_{n_1}(k_\star\mbr_1),
\end{equation}
where $\mrm V_{n_1}$ is defined in \eqref{def:u_n and v_n for a cylinder}. 

The unknowns $k_\star$ and $F_{nn_1}(\lambda_1)$ can be determined by substituting \eqref{eq:fn effective modes decomposition} into \eqref{eq:ensemble wave equation}. The details are shown in Appendix~\ref{sec:effective eigensystem} with the resulting equation being 
\begin{equation}
\label{eq:regular eigensystem (matrix form)}
    F_{nn_1}(\lambda_1) 
    +
    \sum_{n'n_2} \delta_{n_2-n_1+n'-n}\mrm T_n(\lambda_1)\int_\Sc \Nc_{n'-n}^{12}(k,k_\star)F_{n'n_2}(\lambda_2)\mathfrak{n}(\lambda_2)\,\dd\lambda_2 = 0,
\end{equation}
where
\begin{align}
    \label{def:matcal N}
&\Nc_l^{12}(k,k_\star)  := 
    2\pi \frac{\mrm N_l(k a_{12},k_\star a_{12})}{k_\star^2 - k^2}
   -2\pi \int_{a_{12}}^{b_{12}}\mrm J_{l}(k_\star r)\mrm H_{l}(kr)\delta\pc(r)r\dd r,
   \\
   & \mrm N_l(x,y)  := x \mrm H'_{l}(x)\mrm J_{l}(y) - y\mrm H_{l}(x)\mrm J'_{l}(y).
\end{align}


The above is a nonlinear eigenvalue problem, which is why we refer to $k_\star$ as an eigenvalue. After calculating $k_\star$ we can calculate the eigenfunctions $F_{n'n_2}(\lambda_1)$ by solving the linear system \eqref{eq:regular eigensystem (matrix form)}, though in practice it is far better to calculate both $k_\star$ and $F_{n'n_2}(\lambda_1)$ using the modal decomposition as we do in Section~\ref{sec:effective T-matrix}.



\begin{figure}[ht]
\fcolorbox{gray!20}{gray!20}{
\begin{minipage}{.999\linewidth}
    \centering
    \textbf{Multiple effective wavenumbers}\\[5pt]
\includegraphics[scale=.5]{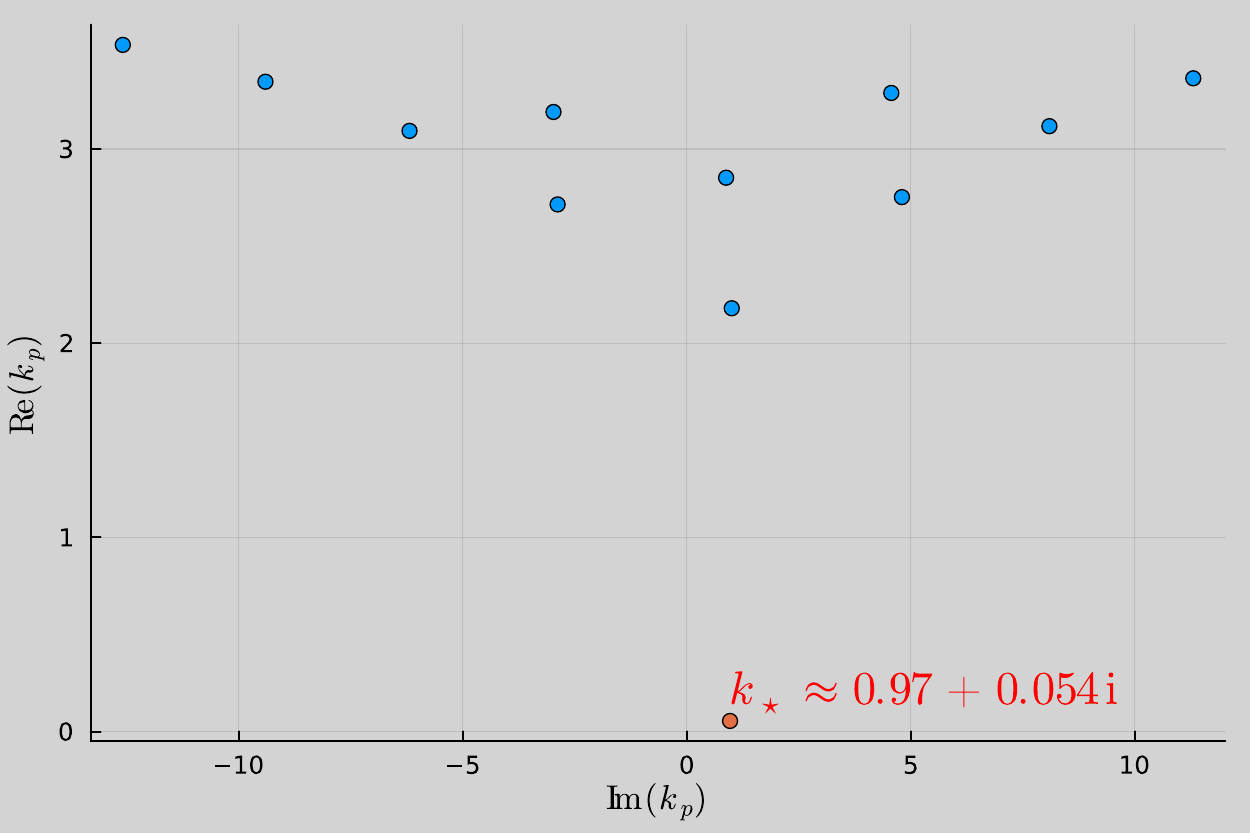}
\caption{Shows 11 of the eigenvalues $k_p$ of \eqref{eqn:dispersion} for sound soft particles of radius $a=1$ confined in a cylinder of radius $R=10$ when $\omega=1$. $k_\star$ corresponds to the one with the smallest imaginary part. The $x$-axis is the real part and $y$-axis is the imaginary part. The other wavenumbers have a much larger imaginary part leading to evanescent waves inside the random material.
\label{fig:effective wavenumbers}
}
\end{minipage}
}
\end{figure}

\subsection{The ensemble boundary condition}

The eigensystem \eqref{eq:regular eigensystem (matrix form)} is not enough to fully determine the $F_{n n_1}$, for instance if $F_{n n_1}$ is a solution to the eigensystem then so is $\alpha F_{n n_1}$ for any scalar $\alpha$. 

To fully determine $F_{n n_1}$ we need to substitute \eqref{eq:fn effective modes decomposition} into the ensemble boundary condition \eqref{eq:ensemble boundary equation}. The details are shown in Appendix \ref{sec:effective boundary}, where we deduce the ensemble boundary conditions \eqref{eq:general ensemble boundary condition} without specifying the region $\Rc_2$, followed by choosing $\Rc_2$ to be a cylinder with radius $R - a_2$ which results in the boundary condition:
\begin{equation}
\label{eq:circular ensemble boundary condition}
    \tcboxmath{
     g_{N}+ \frac{2\pi}{k_\star^2-k^2}\sum_{n' }\int_\Sc 
 F_{n' (N-n')}(\lambda_2) 
    N_{N-n'}(kR_2,k_\star R_2) \mathfrak{n}(\lambda_2)\,\dd\lambda_2= 0,
    }
\end{equation}
where $N_{l}$ is defined by \eqref{def:matcal N}. 

\section{An effective cylinder}

\label{sec:effective T-matrix}

\subsection{Modal decomposition of the problem}

In this section, we assume that the particles are confined in a circular disc $\Rc$ with radius $R$. Since \eqref{eq:average governing equation} is linear with respect to $g_n$, it can be decomposed in simpler and independent equations:  let $\ensem{f_{n,N}}(\mbr_1,\lambda_1)$ be the solution when substituting  $g_n=\delta_{n,N}$ in \eqref{eq:average governing equation}, then
\begin{multline}
\label{eq:average governing equation (modal)}
\ensem{ f_{n,N}}(\mbr_1,\lambda_1)
=   \mrm T_n(\lambda_1)\dsp \mrm V_{N-n}(k\mbr_1)
\\  +
\mrm T_n(\lambda_1)\dsp\sum_{n'} \int_\Sc \mathfrak{n}(\lambda_2) \int_{\Rc_2} \mrm U_{n'-n}(k\mbr_1 - k \mbr_2)\ensem{f_{n',N}}(\mbr_1,\lambda_1) \pc(|\mbr_1- \mbr_2|, \lambda_1, \lambda_2) \,\dd \mbr_2\dd\lambda_2,
\end{multline}
where we can recover the solution for any incident wave by using 
\begin{equation}
\label{eq:general solution as sum of modal solutions}
   \ensem{f_{n}}(\mbr_1,\lambda_1) = \sum_N g_N \ensem{f_{n,N}}(\mbr_1,\lambda_1).
\end{equation}
From the 3D version of these effective equations \cite{gower2021effective} we know that \eqref{eq:regular eigensystem (matrix form)} can be reduced by using symmetry. We show how to do this for the modal decomposition below, with the result being:
\begin{equation}
\label{eq:modal radial symmetry}
\tcboxmath{
\ensem{f_{n,N}}(\mbr_1,\lambda_1) = \alpha_N F_{n,N}(\lambda_1)\mrm V_{N-n}(k_\star\mbr_1)
}
\end{equation}
where $\alpha_N\in\C$ is some amplitude which is introduced for later convenience. Below is the proof of \eqref{eq:modal radial symmetry}.
\vspace{10pt}

\begin{proof}

Using the rotational symmetry of the modal source, we simplify the form of the modal solutions $\ensem{ f_{n,N}}(\mbr_1,\lambda_1)$. To this end, we denote by $\mbR_\phi$ the rotation matrix of angle $\phi$ and replace $\mbr_1$ with $\mbR_\phi\mbr_1$  in \eqref{eq:average governing equation (modal)}:
\begin{multline} \label{eqn:modal integral 1}
\langle f_{n,N}\rangle(\mbR_\phi\mbr_1,\lambda_1)
=\mrm T_n(\lambda_1)\dsp \mrm V_{N-n}(k\mbR_\phi\mbr_1)
\\+\mrm T_n(\lambda_1)\dsp\sum_{n'} \int_\Sc \mathfrak{n}(\lambda_2) \int_{\Rc_2} \mrm U_{n'-n}(k\mbR_\phi\mbr_1-k\mbR_\phi\mbr_2)\ensem{f_{n',N}}(\mbR_\phi\mbr_2,\lambda_2) \pc(|\mbr_1-\mbr_2|) \dd \mbr_2\dd\lambda_2,
\end{multline}
where we changed the integration variable from $\mbr_2$ to $\mbR_\phi\mbr_2$, which is possible for any rotation as $\Rc_2$ is a disc.  Then from~\eqref{def:u_n and v_n for a cylinder} we deduce the property 
\[
\mrm U_n(\mbR_\phi \mbr_1) = \mrm U_n(\mbr_1) \mrm e^{\mrm i n \phi} \quad \text{and} \quad \mrm V_n(\mbR_\phi \mbr_1) = \mrm V_n(\mbr_1) \mrm e^{\mrm i n \phi},
\]
which used in \eqref{eqn:modal integral 1}, and then multiplying both sides of the equation with $\mrm e^{-\mrm i(N-n)\phi}$ leads to
%
\begin{multline}
    \ensem{ f_{n,N}}(\mbR_\phi\mbr_1,\lambda_1)\mrm e^{-\mrm i(N-n)\phi}
    =
    \mrm T_n(\lambda_1)\dsp  \mrm V_{N-n}(k\mbr_1) +
    \\[6pt]
\mrm T_n(\lambda_1)\dsp\sum_{n'} \int_\Sc \mathfrak{n}(\lambda_2) \int_{\Rc_2} \mrm U_{n'-n}(k\mbr_1 - k \mbr_2)\mrm e^{-\mrm i(N-n')\phi}\ensem{f_{n',N}}(\mbR_\phi\mbr_2,\lambda_2) \pc(|\mbr_{1} - \mbr_{2}|) \dd \mbr_2\dd\lambda_2.
\end{multline}
Now note that both $\ensem{ f_{n,N}}(\mbR_\phi\mbr_1,\lambda_1)\mrm e^{-\mrm i(N-n)\phi}$ and $\ensem{f_{n,N}}(\mbr_1)$ solve exactly the same integral equation. So by assuming uniqueness, i.e. that there is only one solution to the above, we conclude that
\begin{equation}
\ensem{f_{n,N}}(\mbr_1)= \ensem{f_{n,N}}(\mbR_\phi\mbr_1) \mrm e^{-\mrm i(N-n)\phi},
\end{equation}
for any $\mbr_1$ and $\phi$. Let $(r_1,\theta_1)$ be the polar coordinates of $\mbr_1$, then, without lose of generality, we then choose $\phi=-\theta_1$ which leads to
\begin{equation}
\ensem{f_{n,N}}(r_1,\theta_1,\lambda_1) = \ensem{f_{n,N}}(r_1,0,\lambda_1)\mrm e^{\mrm i(N-n)\theta_1}.
\end{equation}
Finally, because $\ensem{f_{n,N}}(\mbr_1)$ satisfies a wave equation \eqref{def:effective wavenumber assumption} it can be written in a series of the form \eqref{eq:fn effective modes decomposition}. Using the symmetry above we see that only one term in this series remains shown by \eqref{eq:modal radial symmetry}. 
\end{proof}

\subsection{The modal dispersion equation}

We can deduce a simpler effective eigensystem and dispersion equation by using the symmetry \eqref{eq:modal radial symmetry}.

To start we substitute the expansions \eqref{eq:fnN decomposition} and \eqref{eq:modal radial symmetry} into the modal decomposition~\eqref{eq:general solution as sum of modal solutions} to obtain
\begin{equation}
\sum_{n_1}F_{nn_1}(\lambda_1)\mrm V_{n_1}(k_\star\mbr_1) 
=
\sum_{N_1} \alpha_{N_1} g_{N_1}F_{n,{N_1}}(\lambda_1)\mrm V_{{N_1}-n}(k_\star\mbr_1)
\end{equation}
Then since $\mrm V_n$ form an orthogonal basis of functions,
\begin{equation} \label{eqn:F as modal sum}
    F_{n n_1} = \sum_{N_1} \delta_{{N_1} - n_1 -n} \alpha_{N_1} g_{N_1} F_{n,{N_1}}.
\end{equation}

To simplify the effective eigensystem~\eqref{eq:regular eigensystem (matrix form)} we consider one mode at a time by taking $g_{N_1} = \delta_{{N_1}-N}$, which used in \eqref{eqn:F as modal sum} implies that we can substitute $F_{n n_1}=\delta_{N - n_1 -n} \alpha_N F_{n,N}$ into \eqref{eq:regular eigensystem (matrix form)}. The result after some algebraic manipulations is 
%
\begin{equation}
\label{eq:eigensystem (modal)}
    F_{n,N}(\lambda_1) 
    +
    \sum_{n'} \mrm T_n(\lambda_1)\int_\Sc \Nc_{n'-n}^{12}(k,k_\star)F_{n',N}(\lambda_2)\mathfrak{n}(\lambda_2)\,\dd\lambda_2 = 0.
\end{equation}
The above equation is identical to the case of the eigensystem for plane waves \cite{gower2018reflection}, and matches also the eigensystems for a single type of particle \cite{gower2019multiple,linton2005multiple} when taking $\mathfrak{n}(\lambda) = \delta(\lambda - \lambda_1)$. This result is somewhat expected as the ensemble wave equation \eqref{eq:ensemble wave equation} does not depend on the incident wave and material geometry, which also explains why the modal index $N$ only appears in $F_{n,N}$ in the above.

Instead of solving \eqref{eq:regular eigensystem (matrix form)}, it is far simpler to solve the above, and then write the general solution in the modal form \eqref{eqn:F as modal sum}. In practice, to solve \eqref{eq:eigensystem (modal)}, we can discretise the integral over $\Sc$ as a set of reals $\{t_1,\dots,t_S\}$. Then define a block vector $\mathbf{F}$ containing the entries $ F_{n}(t_s)$ for $n = - M, -M +1, \ldots, M -1, M$, for some finite $M$, and for $s=1,\dots,S$, so that the eigensystem becomes
\begin{equation}
\label{eq:modal eigensystem matrix form}
    (\mathbf{I}+\mathbf{M})\cdot\mathbf{F} = \vec 0
\end{equation}
where 
\begin{equation}
    M_{nn',ss'}(k_\star)= \mrm T_n(t_s) \,\Nc_{n'-n}^{12}(k,k_\star)\,\mathfrak{n}(t_{s'}).
\end{equation}
The parameter $k_\star$ is then obtained by solving the equation 
\begin{equation} \label{eqn:dispersion}
    \mrm{det} [\mathbf{I}+\mathbf{M}(k_\star)] = 0.
\end{equation}

\begin{optionalnote}[label=rem:multiple wavenumbers]{Multiple wavenumbers}
The dispersion equation \eqref{eq:eigensystem (modal)} has infinitely many eigenvalues $k_p$ with $p =1, 2, \ldots $. Consequently, $\ensem{f_n}(\mbr_1)$ can more generally be written as a sum over all the eigenvalues $k_p$ and their corresponding eigenfunctions \cite{gower2019proof}, which leads to more accurate solutions. However, only a small difference in comparison to using just the eigenvalue $k_\star$ with the smallest imaginary part is observed (cf. \cite{gower2019multiple,gower2019proof,gower2021effective} for details). For this reason, and for simplicity, we only account for the one wavenumber $k_\star$ in this paper. See Figure~\ref{fig:effective wavenumbers} for a typical distribution of the many eigenvalues of \eqref{eq:eigensystem (modal)}.
\end{optionalnote}
\subsection{The modal ensemble boundary condition} 

To determine the $\alpha_N$ that appear in \eqref{eqn:F as modal sum} we need to use the ensemble boundary condition. The simplest way to do this is again to take $g_{N_1} = \delta_{N_1 - N}$ in \eqref{eqn:F as modal sum} and then substitute the result into \eqref{eq:circular ensemble boundary condition} to obtain
\begin{equation}
\label{eq:boundary condition}
    \tcboxmath{
    1+ \frac{2\pi \alpha_N}{k_\star^2-k^2}\sum_{n'}\int_\Sc 
  F_{n',N}(\lambda_2) 
    N_{ N-n'}(kR_2,k_\star R_2)\mathfrak{n}(\lambda_2)\,\dd\lambda_2= 0,
    }
\end{equation}
which we can use to determine:
\begin{equation}
\label{def:alpha}
\alpha_N = -\frac{k_\star^2-k^2}{2\pi}\left(
\sum_{n'}\int_\Sc 
 F_{n',N}(\lambda_2) 
    N_{ N-n'}(kR_2,k_\star R_2) \mathfrak{n}(\lambda_2)\,\dd\lambda_2
\right)^{-1}.
\end{equation}

\subsection{The effective T-matrix}
As discussed Section~\ref{sec:effective T-matrix}, the effective T-matrix $\mathcal T_{n,N}$ can easily describe the average scattered wave for any incident wave through:
\begin{equation}
\label{eq:T-matrix relation}
\ensem{\mathfrak{F}_n} = \sum_{N=-\infty}^{+\infty}\Tc_{n,N} g_N,
\end{equation}
where the $\ensem{\mathfrak{F}_n}$, given by \eqref{eq:averaged scattering coefficients}, are the average coefficients of the waves scattered from the whole cylinder $\Rc$. Note the above holds for any choice of $g_N$ and $\Tc_{n,N}$ does not depend on $g_N$.


To calculate $\mathcal T_{n,N}$ we substitute the modal decomposition \eqref{eq:general solution as sum of modal solutions} into \eqref{eq:averaged scattering coefficients} to obtain
\begin{equation}
    \ensem{\mathfrak{F}_n} = \sum_Ng_N\int_\Sc\mathfrak{n}(\lambda_1)\int_{\Rc_1}\sum_{n'}\mrm V_{n'-n}(-k\mbr_1)\ensem{f_{n',N}}(\mbr_1,\lambda_1)\,\dd\mbr_1\dd\lambda_1.
\end{equation}
Comparing the above with the definition of the effective T-matrix \eqref{eq:T-matrix relation}, it is clear that 
\begin{equation}
\label{eq:effective T-matrix (exact formula)}
\Tc_{n,N} = 
\int_\Sc\mathfrak{n}(\lambda_1)\int_{\Rc_1}\sum_{n'}\mrm V_{n'-n}(-k\mbr_1)\ensem{f_{n',N}}(\mbr_1,\lambda_1)\,\dd\mbr_1\dd\lambda_1.
\end{equation}
To calculate the above, we follow the same steps shown in Appendix \ref{app:effective method}. Specifically, use use Green's second identity \eqref{eqn:Greens 2nd identity}, the regular expansion  \eqref{eq:modal radial symmetry}, and the orthogonality of the $\mrm V_n$ functions to conclude that

\begin{equation}
   \Tc_{n,N} = 
\delta_{N-n} \frac{2\pi\alpha_n}{k_\star^2-k^2}\int_\Sc\mathfrak{n}(\lambda_1)\sum_{n'}  F_{n',n}(\lambda_1)\Qc_{n-n'}(kR_1,k_\star R_1) \dd \lambda_1
\end{equation}
where $R_1$ is the radius of the disc $\Rc_1$ and we defined:
\begin{equation}
    \Qc_l(x,y):= x \mrm J_{l}'(x) \mrm J_l(y) - y\mrm J_{l}(x) \mrm J_{l}'(y).
\end{equation}
Finally, we substitute $\alpha_n$ given by \eqref{def:alpha}, which results in
\begin{equation} \label{eq:effective T-matrix}
   \Tc_{n,N} =  -
\delta_{N-n} \frac{\dsp \int_\Sc\mathfrak{n}(\lambda_1)\sum_{n'}  F_{n',n}(\lambda_1)\Qc_{n-n'}(kR_1,k_\star R_1) \dd \lambda_1}{ \dsp
\int_\Sc  \mathfrak{n}(\lambda_1)
 \sum_{n'} F_{n',n}(\lambda_1) 
    N_{n-n'}(kR_1,k_\star R_1) \,\dd\lambda_1}\cdot
\end{equation}


Note that once the effective T-matrix is known, the scattering from any incident field can be computed with \eqref{eq:averaged scattered field (circular region)} after decomposing the incident in modes \eqref{eq:incident field intro}. For example, in Figure \ref{fig:pressure field},  we have plotted the total pressure field $u:=\ui+\us$ resulting from an incident plane wave and a point source.

\begin{figure}[ht]
\fcolorbox{gray!20}{gray!20}{
\begin{minipage}{.999\linewidth}
    \centering
    \textbf{Average field for two different sources}\\[5pt]
\includegraphics[scale=.4,trim={2cm 0cm 2cm 0cm},clip]{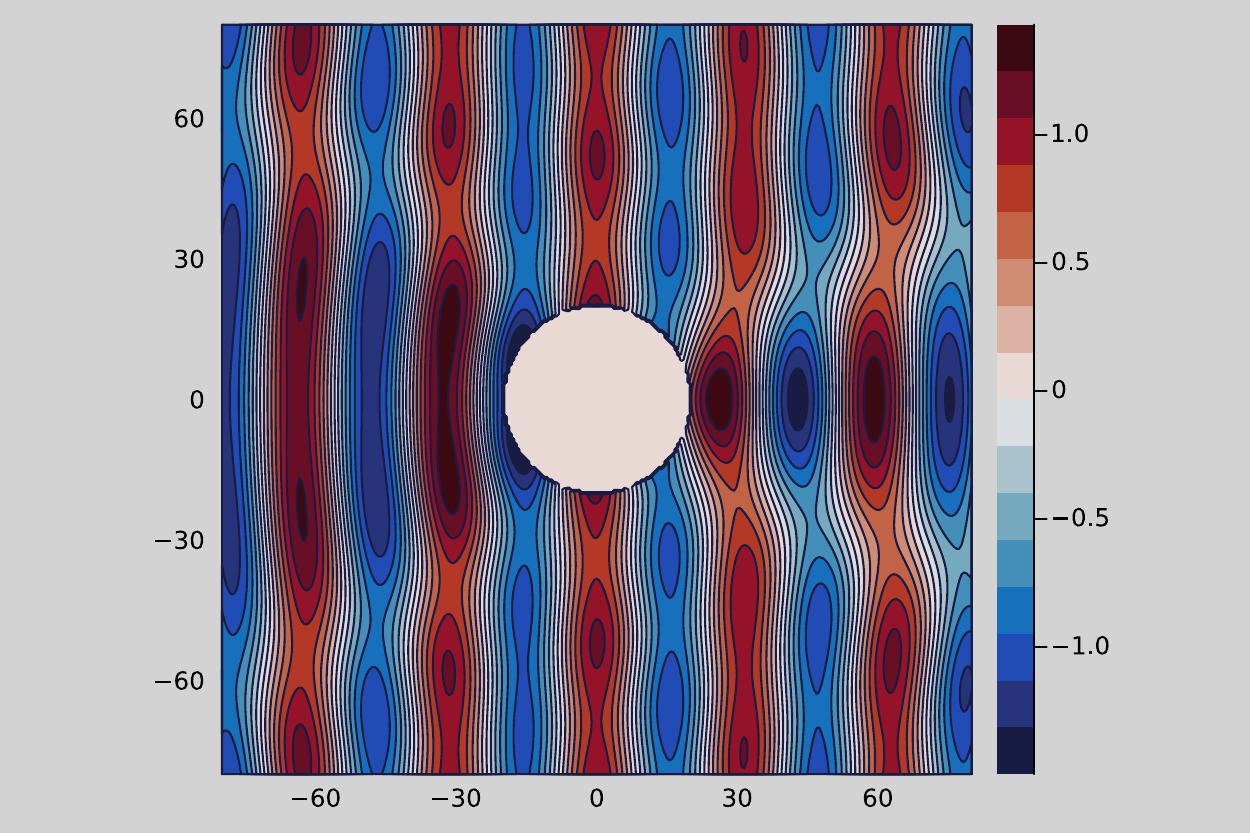} \hspace{10pt}
\includegraphics[scale=.4,trim={2cm 0cm 2cm 0cm},clip]{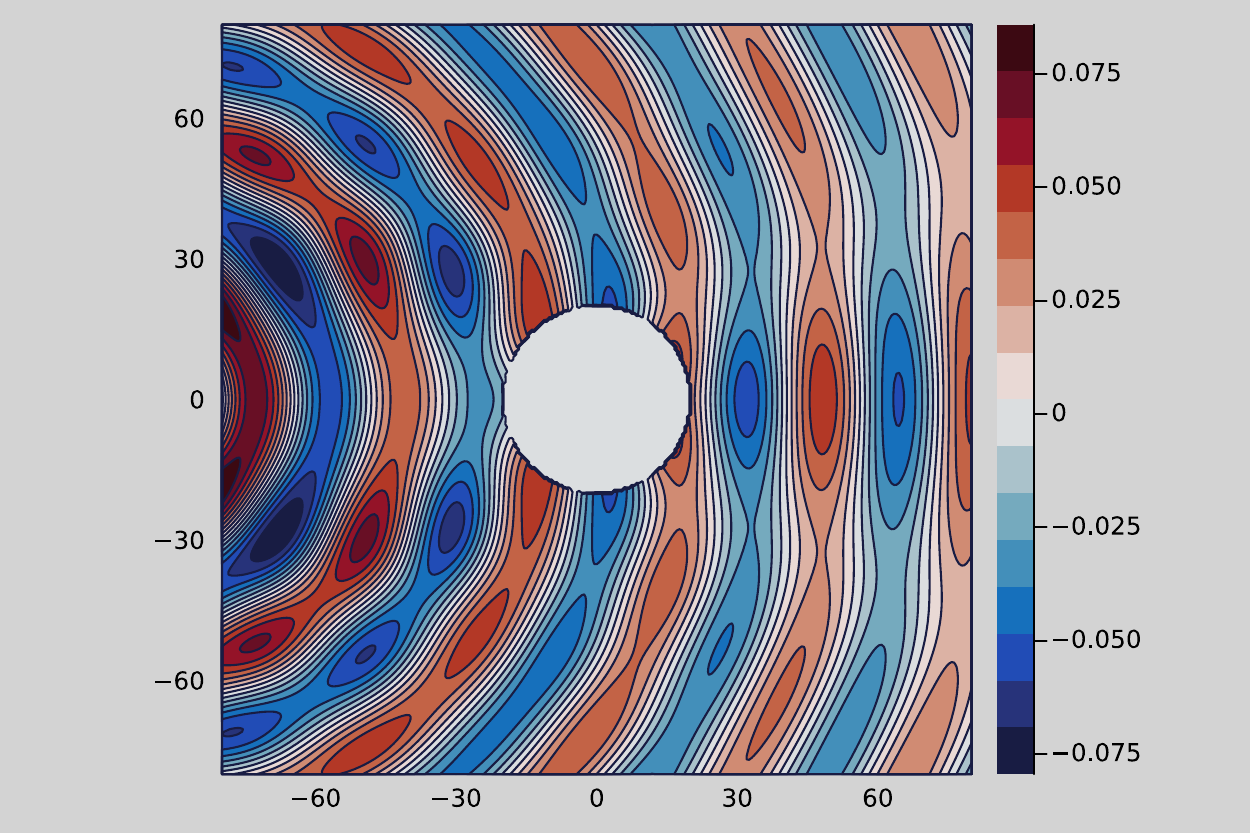}
\caption{Average pressure field when the incident field is a plane wave (left) and a point source (right). The material is made of sound-hard particles of radius one confined in a disc of radius 10.
\label{fig:pressure field}
}
\label{fig:plane wave scattering}
\end{minipage}\hfill
}
\end{figure}

\subsection{Monopole particles only}

\label{sec:kstar interpretation}

The effective T-matrix \eqref{eq:effective T-matrix} resembles the T-matrix for a homogeneous cylinder, see for example \eqref{eq:T-matrix-acoustics}. In fact, it is a weighted average of the factors of a homogeneous T-matrix, as explained in the introduction. From this observation we see that if the particles are monopole scatterers we obtain a significant simplification.   

Let us assume here that the particles scatter only monopole waves, in which case the scattered field~\eqref{eq:scattered field intro} becomes
\begin{equation}
\us(\mbr) = \sum_{i=1}^Jf_0^i\mrm H_0(kr-kr_i),
\end{equation} 
which leads to $\ensem{f_{n,N}}(\mbr_1)= 0$ if $n\neq0$ and, as a result of \eqref{eq:modal radial symmetry}, $F_{n,N}=0$ if $n \neq 0$.  Substituting this result into the effective T-matrix \eqref{eq:effective T-matrix}, and assuming that the radius $R_1$ of the region $\Rc_1$ is the same for every type of particle $\lambda_1$, we obtain:
\begin{equation}
\label{eq:effective T-matrix (monopole)}
\tcboxmath{
     \Tc_n^\mrm M = 
    - \frac{
  \dsp
k\mrm J'_{n}(k\tilde{R})\mrm J_{n}(k_\star\tilde{R})-k_\star\mrm J_{n}(k\tilde{R})\mrm J'_{n}(k_\star\tilde{R})
  }
  {
  \dsp
k\mrm H'_{n}(k\tilde{R})\mrm J_{n}(k_\star\tilde{R})-k_\star\mrm H_{n}(k\tilde{R})\mrm J'_{n}(k_\star\tilde{R}) 
  } 
  \quad\text{(monopole scatterers).}
  }
\end{equation}
This corresponds to the T-matrix of a homogeneous cylinder of radius $\tilde{R}=R-a$, sound speed $c_\star=\omega/k_\star$ and mass density $\rho_\star=\rho$ where $\rho$ is the density of the host medium (cf.  \eqref{eq:T-matrix-acoustics}).

\section{Numerical results}

The numerical results obtained in this paper use the open-source package EffectiveTMatrix.jl written in Julia \cite{EffectiveTMatrix}. More details on the simulations below can be found on the GitHub page \url{https://github.com/Kevish-Napal/EffectiveTMatrix.jl}.

\subsection{Optimized Monte Carlo simulations}

We use a Monte Carlo method to validate our theoretical results \eqref{eq:effective T-matrix} and \eqref{eq:effective T-matrix (monopole)}. To develop an efficient Monte Carlo method we rely on the following symmetry of the modes:
\begin{equation}
\label{eq:modal scattering}
    \ui(\mbr)=\mrm J_N(kr) \mrm e^{\mrm iN\theta}
    \implies
    \ensem{\us}(\mbr)=\mrm T_N \mrm H_N(kr) \mrm e^{\mrm iN\theta}.
\end{equation}
This result is easily obtained from \eqref{eq:averaged scattered field (circular region)} with the specific choice $g_n=\delta_{n,N}$, which substituted into \eqref{def:effective T-matrix} leads
\begin{equation}
\label{eq:modal MC}
    \mrm T_N = \ensem{\mathfrak{F}_{N}}.
\end{equation}
 In other words, $\mrm T_N$ can be numerically estimated by simulating the waves scattered from one particle configuration at a time by using \eqref{eq:f_n^i governing equation}, and then taking the average of $\mathfrak F_N$ \eqref{def:mathfrak Fn} over many different particles configurations. 
\vspace{10pt}

To illustrate the efficiency of this Monte Carlo method  we compare it with another method which directly computes $\ensem{\us}$, which has been commonly used in the literature \cite{kong2004scattering}. For this second method we use \eqref{eq:scattered field intro} to compute $\ensem{\us}(R,0)$. Then from \eqref{eq:modal scattering} we can also compute $\mrm T_N$ with  
\begin{equation}
\label{eq:naive MC}
    \mrm T_N =  \mrm H_N^{-1}(kR)\ensem{\us}(R,0).
\end{equation}

The two methods \eqref{eq:modal MC} and \eqref{eq:naive MC} are compared in Figure \ref{fig:histogram}. The standard deviation of the mean of the second method is larger than the first one, resulting in a slower convergence. The reason is that \eqref{eq:naive MC}, in contrary to \eqref{eq:modal MC}, includes all the modes of each scattered field computed for a specific particles configuration:
\begin{equation}
    \ensem{\us}(R,0) =\bigl \langle \sum_{n} \mathfrak{F}_{n}\mrm H_n(kR) \bigr \rangle.
\end{equation}
While the terms  $n\neq N$ of the  sum vanish on average, they significantly contribute to the standard deviation of the mean in \eqref{eq:naive MC}.

\begin{figure}[ht]
\fcolorbox{gray!20}{gray!20}{
\begin{minipage}{.999\linewidth}
    \centering
    \textbf{Optimized Monte Carlo simulations}\\[5pt]
\includegraphics[scale=.4]{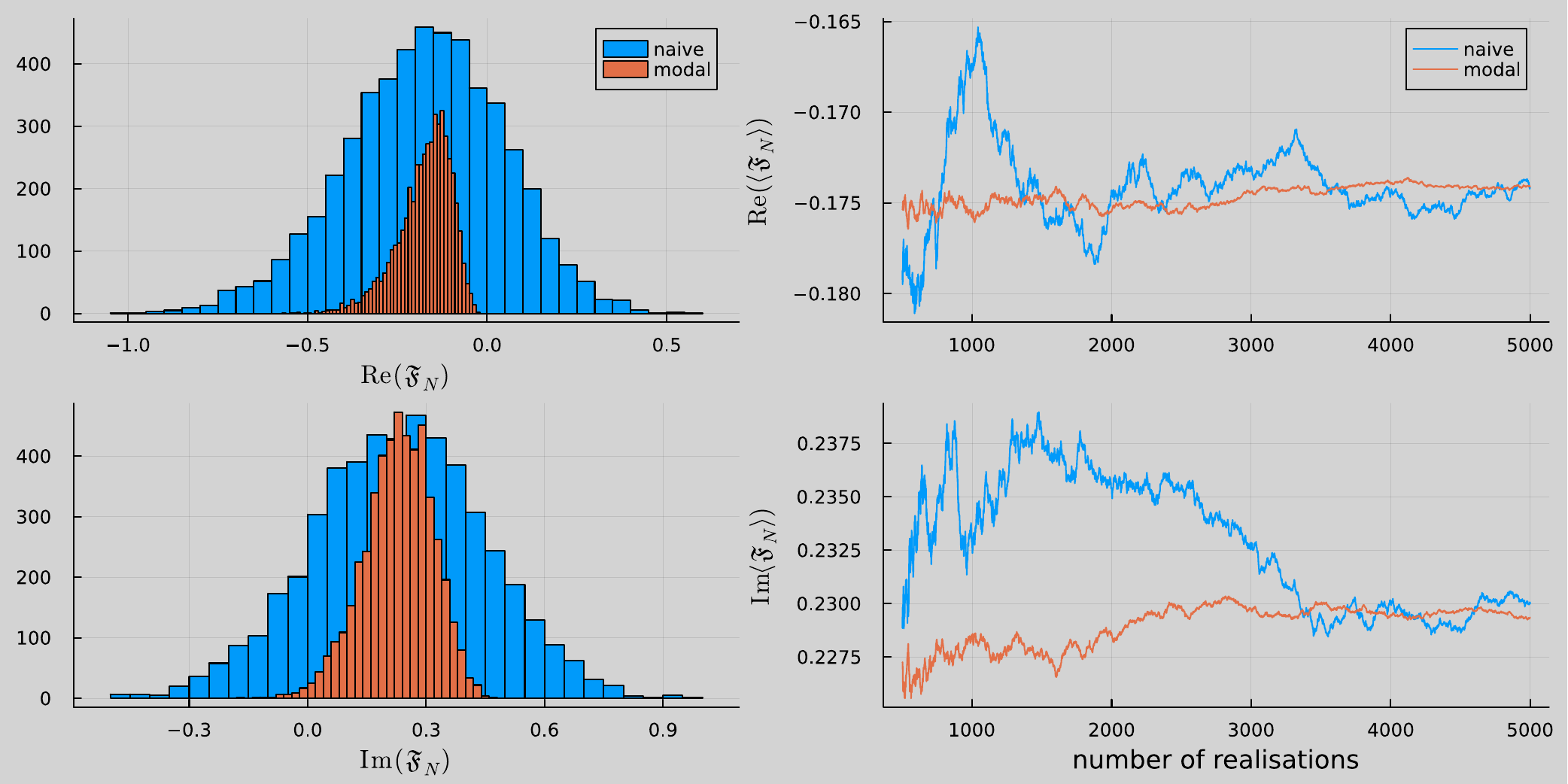}
\caption{Computation of $\mrm T_0$ using the two different methods \eqref{eq:modal MC} (modal) and \eqref{eq:naive MC} (naive).  We computed 5000 realisations of these quantities for different configurations of particles, the plots on the top and bottom respectively correspond to the real and imaginary parts of the results. The distribution of the results is reported on the histograms on the left. The plots on the right correspond to the cumulative average of the realisations. Both methods converge to the same limit, however, the modal method converges faster and presents  a lower standard deviation of the mean.
\label{fig:histogram}
}
\end{minipage}\hfill
}
\end{figure}

\subsection{Validation of the Effective Waves Method}

We validate the Effective Waves method \eqref{eq:effective T-matrix (exact formula)} against the Monte Carlo method \eqref{eq:modal MC} for several frequencies $\omega$ ranging in $\Omega:=[0.05:0.015:1.5]$. To this end we define the relative error, averaged over frequencies:  
\begin{equation}
\label{def:relative errors}
    \epsilon_n := \frac{1}{|\Omega|}\sum_{\omega\in\Omega}\frac{|\mrm T_n^\text{MC}(\omega) - \mrm T_n^\text{EWM}(\omega)|}{|\mrm T_n^\text{MC}(\omega)|}
\end{equation}
where $\mrm T_n^\text{MC}(\omega)$ is obtained following the Monte Carlo method \eqref{eq:modal MC} and $\mrm T_n^\text{EWM}(\omega)$ following the Effective Waves method \eqref{eq:effective T-matrix (exact formula)}. A few plots of $\mrm T_n^\text{MC}(\omega)$ and $\mrm T_n^\text{EWM}(\omega)$ are provided in Figures \ref{fig:intro MC result} and \ref{fig:Dirichlet MC result}. The values of $\epsilon_0$, $\epsilon_1$, $\epsilon_2$, $\epsilon_3$, $\epsilon_4$ for the cases of sound soft and sound hard particles are reported in Table \ref{tab:relative errors}.

\begin{figure}[ht]
\fcolorbox{gray!20}{gray!20}{
\begin{minipage}{.999\linewidth}
    \centering
    \textbf{Monte Carlo results: sound-soft particles}\\[5pt]
\includegraphics[scale=.55,trim={0cm 0cm 10.5cm 2.5cm},clip]{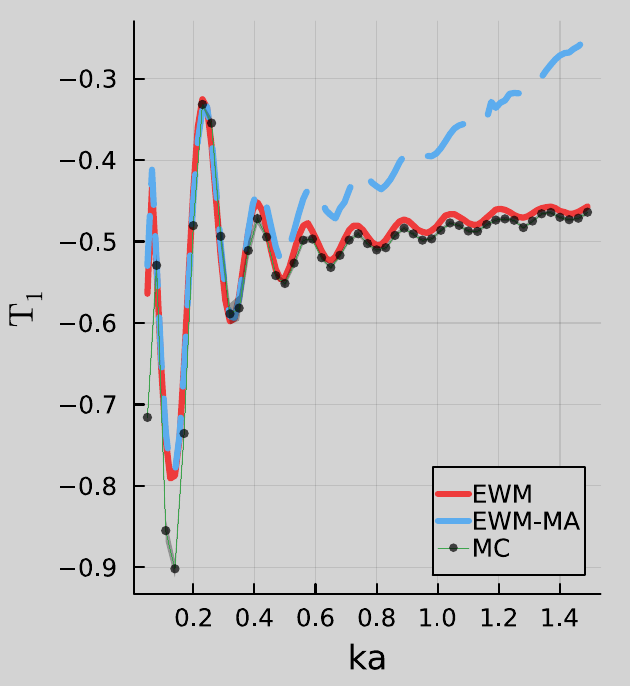}
\includegraphics[scale=.55,trim={0cm 0cm 10.5cm 2.5cm},clip]{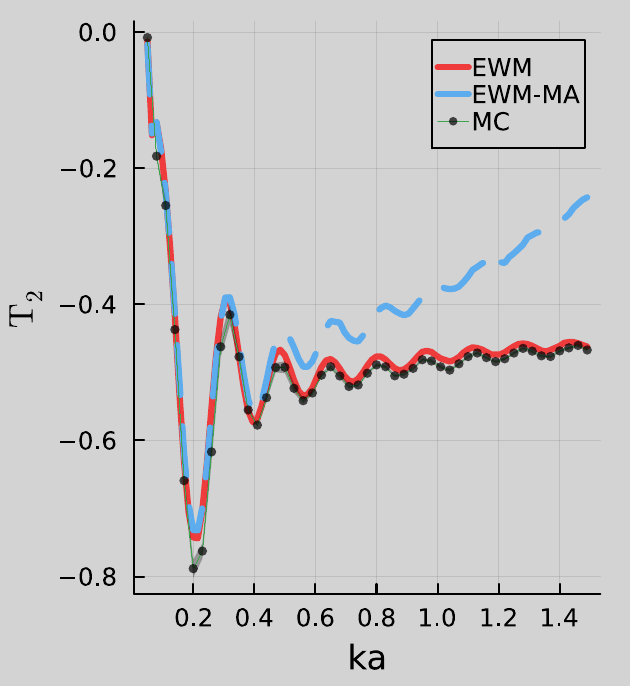}
\includegraphics[scale=.55,trim={0cm 0cm 10.5cm 2.5cm},clip]{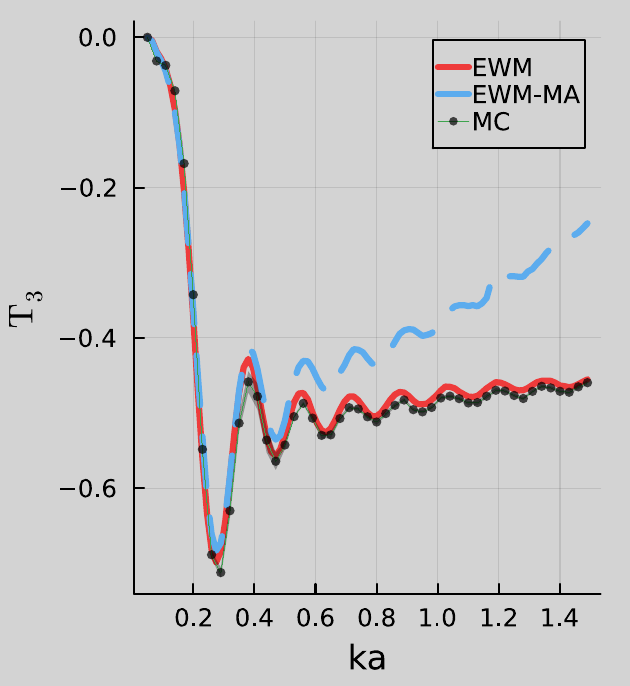}
\includegraphics[scale=.55,trim={0cm 0cm 10.5cm 2.5cm},clip]{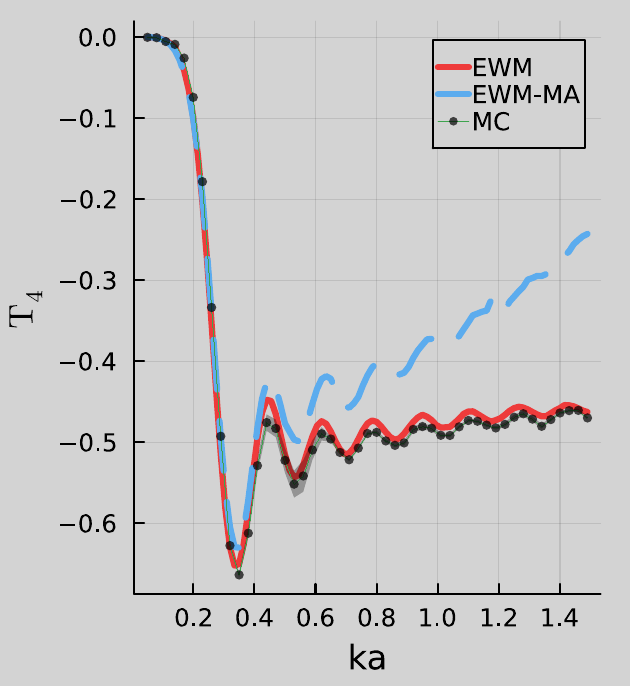}
\caption{Compares various methods to calculate the components $\mrm T_{1}$, $\mrm T_{2}$, $\mrm T_{3}$ and $\mrm T_{4}$  of the T-matrix of a cylinder filled with sound soft particles. The solid red line is our effective waves method (EWM) \eqref{eq:effective T-matrix}, the black points are from the Monte Carlo method (MC) \eqref{eq:modal MC}, and the dashed blue line is our method when only monopole scattering is accounted for (EWM-MA) \eqref{eq:effective T-matrix (monopole)}. The general expression of the effective T-matrix matches the Monte Carlo results. The EWM-MA method only matches well with the Monte Carlo for low frequencies.
\label{fig:Dirichlet MC result}
}
\end{minipage}\hfill
}
\end{figure}

\begin{table}[ht]
\centering
\begin{tabular}{|l|c|c|c|c|c|}
\cline{2-6}
\multicolumn{1}{c|}{} & $\epsilon_0$ & $\epsilon_1$ &$\epsilon_2$ &$\epsilon_3$ &$\epsilon_4$ \\
\hline
sound soft particles & $6.75 \mrm e^{-2}$  & $4.59\mrm  e^{-2}$ & $5.40 \mrm e^{-2}$ & $6.45 \mrm e^{-2}$ & $7.45 \mrm e^{-2}$ \\ \hline
sound hard particles & $2.66 \mrm e^{-2}$  & $2.43 \mrm e^{-2}$ & $2.33 \mrm e^{-2}$ & $2.35 \mrm e^{-2}$ & $2.37 \mrm e^{-2}$ \\ \hline
\end{tabular}
\caption{Relative errors $\epsilon_n$ defined by \eqref{def:relative errors}  in the cases of sound soft and sound hard. The particles are of radius 1 and confined in a circular area of radius 20. The volume fraction is 0.05 which corresponds to 19.7 particles on average. 
\label{tab:relative errors}}
\end{table}

\section{Conclusion}
\textbf{Main goal.}
Our main goal was to describe how an incident wave is scattered from a cylinder filled with smaller cylinders, which we have called particles, that are placed in a disordered but correlated way. We describe this correlation through the inter-particle pair correlation, see \eqref{def:pair correlation}. The literature so far has focused on plane waves scattered from a halfspace or plate filled with a particulate \cite{karnezis2023average}. There has been at least one paper on solving this scenario, but used an ad hoc method, whereas here everything is deduced from first principles making only two assumptions: such as the Quasi-Crystalline Approximation (QCA) \cite{lax1952multiple}, and expressing the average field as a sum of effective waves, which has been shown to be the analytic solution \cite{gower2019proof}. One of the key advantages of describing the scattering from a cylinder with 2D particles is that it is far easier to validate this scenario with direct numerical simulations. Validation is still necessary as the theory requires the use of QCA whose validity is not clearly established. Further, validating this scenario also serves to validate the predicted effective wavenumbers from any material geometry \cite{gower2021effective}.


\textbf{Modal scattering.} A method that allowed us to greatly simplify both the theoretical and Monte Carlo calculations which we use to validate this work, was to make use of all symmetries present. We achieved this by solving for each polar mode of the incident wave separately. This simple, but effective technique, leads us to an effective T-matrix given by equation \eqref{eq:effective T-matrix} that can be used to calculate the average scattered wave from any incident wave, see section \ref{sec:effective waves method} for a brief overview. We note that we were able to describe the scattered field without calculating the average transmitted field. In future work, this may be interesting to do, for example, to clearly identify when different effective wavenumbers are excited \cite{karnezis2023average}.


\textbf{Effective T-matrix.} The result of our theoretical work is summarised by the T-matrix \eqref{eq:effective T-matrix}. Beyond using just QCA, to calculate this T-matrix, we also assumed that only one effective wavenumber $k_\star$ is excited. This is true for a wide range of parameters but it is not always the case. In particular it appears that very strong scattering at moderate frequencies can trigger more than one effective wavenumber to be excited \cite{karnezis2023average,gower2019multiple,gower2021effective}. One possible extension to our work is to include the effect of more than one effective wavenumber.

\textbf{Monopole Scatterers.} One surprising result, is that if the particles only scatter monopole waves, then the effective T-matrix greatly simplifies and becomes \eqref{eq:effective T-matrix (monopole)}. This form is exactly the same as the T-matrix for a homogenous cylinder, one with constant material parameters. We hypothesise that any material filled with monopole scatterers would, on average, respond like a homogeneous material. Monopole scatterers are a good approximation for many types of resonant particles \cite{cotterill2022deeply}. Figure \ref{fig:Dirichlet MC result} compares the results for the monopole scattering approximation with Monte Carlo results for sound-soft particles, which does not assume the particles scatter like monopoles.     


\textbf{Monte Carlo.} Beyond deducing an effective T-matrix for a particulate cylinder, we also developed an efficient Monte-Carlo method, which matched our theoretical predictions very accurately see Figures \ref{fig:intro MC result} and \ref{fig:Dirichlet MC result}. Our numerical validation was for a broad frequency range, but we did not cover a broad range of parameters. Doing this would be valuable future work, and could help clearly identify the limits of QCA and different approximations for the inter-particle pair-correlation.    


\textbf{A prototype for new materials.} The setting we deduce is the ideal case to test new disordered particulates. That is, to use exotic inter-particle pair-correlation to achieve effects such as band-gaps, or impedance matching. The scenario of a cylinder filled with 2D particles is ideal for testing new types of particles, and inter-particle pair-correlations, because they can be easily validated with Monte-Carlo methods. When designing new materials with exotic responses, and stretching the limits of the theory, we need to have a way to validate those predictions. And this paper provides that.



\break 

\begin{appendices}
\section{Bessel functions and translation matrices}

Given two points $\mbx,\,\mby\in\R^2$, we have the following identities where $\mbd =\mbx-\mby$
\begin{equation}
\label{eq:translation identities}
\begin{array}{|lrcll}
\text{i)} & \mathrm V_n(\mby)& = & \dsp\sum_{n'=-\infty}^{+\infty}\mrm V_{n-n'}(\mbd)\mathrm V_{n'}(\mbx), & 
 \text{for all } \mbx,\mbd\in \R^2\\
  \text{ii)} & \mathrm U_n(\mby)& = & \dsp\sum_{n'=-\infty}^{+\infty}\mrm V_{n-n'}(\mbd)\mathrm U_{n'}(\mbx), &
 \text{for all } |\mbx| > |\mbd|\\
 \text{iii)} &  \mathrm U_n(\mby)& = & \dsp\sum_{n'=-\infty}^{+\infty}\mrm U_{n-n'}(\mbd)\mathrm V_{n'}(\mbx), &
 \text{for all } |\mbx| < |\mbd|.
\end{array}
\end{equation}
The above formulas are direct consequences of Graf's theorem (see \cite[Th. 2.11-2.12]{martin2006multiple} for instance).

\section{Ensemble averaging}
\label{app:ensemble_average}

Here we give a brief overview of ensemble averaging so that this paper is more self-contained. For more details, see \cite{gower2021effective,martin2006multiple} and the references within. To simplify computations, we represent one particle configurations with:
\begin{align*}
    & \Lambda = \mbr_1,\lambda_1\dots\mbr_{ J},\lambda_{ J},
    \quad \quad 
    \Lambda^{(1)} = \mbr_2,\lambda_2\dots\mbr_{ J},\lambda_{J},
    \\
   & \Lambda^{(1,j)} = \mbr_2,\lambda_2,\dots,\mbr_{j-1},\lambda_{j-1},\mbr_{j+1},\lambda_{j+1},\dots,\mbr_{ J},\lambda_{ J}.
\end{align*}
Using these definitions, we define the ensemble average, and conditional ensemble averages, of a quantity $\mrm A$, which can depend on the positions and properties of all the particles, as
\begin{align}\label{def:ensemble average}
	& \ensem{\mrm A} :=\int 
	\mrm A(\Lambda)\,\proba(\Lambda)
	\dd\Lambda
 \qquad
 	\ensem{\mrm A}(\mbr_i,\lambda_i):=\int 
	\mrm A(\Lambda)\,\proba(\Lambda^{(i)}|\mbr_i\lambda_i)\dd\Lambda^{(i)},
 \\ \label{def:ensemble average conditional}
 & \ensem{\mrm A}(\mbr_i,\lambda_i; \mbr_j,\lambda_j) :=
 \int \mrm A(\Lambda)\,\proba(\Lambda^{(i,j)}|\mbr_i,\lambda_i; \mbr_j,\lambda_j)\dd\Lambda^{(i,j)},
\end{align}
where the domain of integration for $\Lambda$ is over all possible particles positions and properties. The term $\ensem{\mrm A}(\mbr_i,\lambda_i; \mbr_j,\lambda_j)$ is the ensemble average of $\mrm A$ conditional to  $(\mbr_1,\lambda_1,\mbr_2,\lambda_2)$.
%

Taking an ensemble average \eqref{def:ensemble average} on both sides of \eqref{eq:us compact expression} leads to 
\begin{equation} \label{eqn:compute <Fn>}
\begin{aligned}
\ensem{\mathfrak{F}_n} 
&=
\sum_{i=1}^J\sum_{n'=-\infty}^{+\infty}\int (-1)^{n-n'}\mrm V_{n-n'}(k\mbr_i)\ensem{f_{n'}^i}
(\mbr_i,\lambda_i)\proba(\mbr_i,\lambda_i)\,\dd\mbr_i \dd \lambda_i
\\
&=
J\sum_{n'=-\infty}^{+\infty}\int (-1)^{n-n'}\mrm V_{n-n'}(k\mbr_i)\ensem{f_{n'}^1}(\mbr_1,\lambda_1)\proba(\mbr_1, \lambda_1)\,\dd\mbr_1\dd \lambda_1
\end{aligned}
\end{equation}
where we used the definition of conditional probability: 
$
\proba(\Lambda) = \proba(\mbr_i, \lambda_i)\proba(\Lambda^{(i)})
$,
the definition of conditional average \eqref{def:ensemble average conditional} to introduce the term $\ensem{f_{n'}^i}(\mbr_i, \lambda_i)$, and then used that particles are indistinguishable\footnote{Said in another way, the variables of integration $\mbr_i$ and $\lambda_i$ are just dummy variables which can be all changed to $\mbr_1$ and $\lambda_1$.}. Finally, using \eqref{eq:particle position uniform distribution} and \eqref{def:mathfrak n} leads to formula \eqref{eq:averaged scattering coefficients}.

\subsection{Average governing equation}
\label{app:average governing equation computation}

Here we briefly show how to reach the averaging governing equation by using just one assumption, the Quasi-Crystalline Approximation (QCA). For more details see \cite{gower2021effective}.

Taking the conditional average \eqref{def:ensemble average}${}_2$ of \eqref{eq:f_n^i governing equation} with $i=1$ gives
\begin{multline}
\langle f_n^1\rangle(\mbr_1,\lambda_1)
=
\mrm T_n(\lambda_1)\dsp \sum_{n'}\mrm V_{n'-n}(k\mbr_1)g_{n'}
\\
+\mrm T_n(\lambda_1)\dsp\sum_{j\neq 1}  \sum_{n'} \int\mrm U_{n'-n}(k\mbr_1 -k\mbr_j)f_{n'}^j(\mbr_1,\lambda_1) \proba(\Lambda^{(1)}|\mbr_1,\lambda_1) \dd \Lambda^{(1)}.
\end{multline}
We can simplify the above by using the definition of the conditional average \eqref{def:ensemble average conditional} of $f_{n'}^j(\mbr_1,\lambda_1;\mbr_j,\lambda_j)$, using \eqref{eq:conditional probability}, and that particles are indistinguishable to obtain
\begin{multline}
\label{eq:pre QCA average governing equation}
\langle f_n^1\rangle(\mbr_1,\lambda_1)
=  \mrm T_n(\lambda_1)\dsp \sum_{n'}\mrm V_{n'-n}(k\mbr_1)g_{n'}
\\
+\mrm T_n(\lambda_1)\dsp\sum_{n'} \int_\Sc \mathfrak{n}(\lambda_2) \int_{\Rc_2} \mrm U_{n'-n}(k\mbr_1 - k \mbr_2)\ensem{f_{n'}^2}(\mbr_2,\lambda_2, \mbr_1,\lambda_1) \pc(\mbr_1, \lambda_1; \mbr_2, \lambda_2) \dd \mbr_2\dd\lambda_2.
\end{multline}
%
%
However, this equation is not a closed form equation for $\ensem{f_n^1}(\mbr_1,\lambda_1)$ and an extra assumption is required to proceed further.

 A standard solution found in the literature to tackle the problem mentioned above is to use the quasi-crystalline approximation  (QCA). It is stated as follows:
\begin{equation}
\label{def:QCA}
\ensem{f_n^2}(\mbr_1,\lambda_1;\mbr_2,\lambda_2)\approx\langle f_n^2 \rangle(\mbr_2,\lambda_2),  \quad |\mbr_1 - \mbr_2| \geq a_{12} \quad \text{(QCA).}
\end{equation}
See \cite{gower2018reflection} for a brief discussion on this approximation. 

Finally we use that particles are indistinguishable which implies that $\ensem{f_n^1}(\mbr_1,\lambda_1) = \ensem{f_n^2}(\mbr_2,\lambda_2)$ when $\mbr_1 = \mbr_2$ and $\lambda_1 = \lambda_2$ to substitute 
\begin{equation} \label{eqns:indistinguishable}
    \ensem{f_n^2}(\mbr_2,\lambda_2) = \ensem{f_n}(\mbr_2,\lambda_2)
\quad \text{and} \quad 
\ensem{f_n^1}(\mbr_1,\lambda_1) = \ensem{f_n}(\mbr_1,\lambda_1)
\end{equation}
into \eqref{eq:pre QCA average governing equation}, which together with QCA~\eqref{def:QCA} leads to the average governing equation \eqref{eq:average governing equation}.

\section{The effective waves method}
\label{app:effective method}

\subsection{Derivation of two ensemble equations}
\label{app:derivation of EWE and EBC}

Here we show how to use the effective wave assumption to rewrite the governing equation~\eqref{eq:average governing equation (modal)} into two separate equations: the effective wave equation and the effective boundary conditions. To achieve this we define using set builder notation: 
\[
D(\vec y,a) := \{ \vec x \in \mathcal R : \; |\vec x - \vec y| \leq a \}.
\]

Using the decomposition of the pair correlation function \eqref{eqn:split-pc}, we split the domain of integration in the governing equation \eqref{eq:average governing equation} into two integrals: one over $D(\mbr_1,a_{12})$, another one over $\Rc_2\setminus D(\mbr_1,a_{12})$ in the form
\begin{equation}
\label{eq:averaged governing equation decomposed}
\begin{aligned}
\ensem{&f_{n}}(\mbr_1,\lambda_1)
=   \mrm T_n(\lambda_1) \sum_{n'} \mrm V_{N-n}(k\mbr_1) g_{n'} +
\\[1pt] 
&\hspace{0pt}\mrm T_n(\lambda_1)\dsp\sum_{n'}
\int_\Sc \mathfrak{n}(\lambda_2) 
\int_{\Rc_2\setminus D(\mbr_1,a_{12})} \mrm U_{n'-n}(k\mbr_1-k\mbr_2)\ensem{f_{n'}}(\mbr_2,\lambda_2) \dd \mbr_2\dd\lambda_2 +
\\[1pt] 
&\hspace{0pt}\mrm T_n(\lambda_1)\dsp\sum_{n'} 
\int_\Sc \mathfrak{n}(\lambda_2) 
\int_{D(\mbr_1,a_{12},b_{12})} \mrm U_{n'-n}(k\mbr_1-k\mbr_2)\ensem{f_{n'}}(\mbr_2,\lambda_2) \delta\pc(|\mbr_1-\mbr_2|,\lambda_1,\lambda_2)\dd \mbr_2\dd\lambda_2.
\end{aligned}
\end{equation}
where $D(\mbr_1,a_{12},b_{12}):=D(\mbr_1,b_{12})\setminus D(\mbr_1,a_{12})$ and the annulus  $D(\mbr_1,a_{12},b_{12})$ is completely contain within $\Rc_2$ when 
\begin{equation} \label{eqn:r1 distance}
\mathrm{dist}(\mbr_1, \partial \Rc_2) \geq b_{12}.    
\end{equation}
In this section, and in the whole paper, we only solve~\eqref{eq:average governing equation (modal)} for $\mbr_1$ that satisfies the above. This avoids the boundary layer \cite{gower2019multiple,gower2019proof} which greatly complicates the solution and is only needed when there is a large particle volume fraction, moderate frequencies, and strongly scattering particles.  


The last integral in \eqref{eq:averaged governing equation decomposed} can be simplified by changing the variable of integration to $\mbr = \mbr_2 - \mbr_1$ which leads to the integral
\begin{equation}
\label{def:matcal K}
    \Kc_{n'n}(\mbr_1,\lambda_2) 
    := 
    \int_{D(\vec 0,a_{12},b_{12})} \mrm U_{n'-n}( -k\mbr)\ensem{f_{n'}}(\mbr + \mbr_1,\lambda_2) \delta\pc(r,\lambda_1,\lambda_2)\, \dd \mbr.
\end{equation}

The first integral over $\mbr_2$ in \eqref{eq:averaged governing equation decomposed} can be further simplified by using Green's theorem  to replace the volume integral over $\Rc_2\setminus D(\mbr_1,a_{12})$ by surface integrals: given any two function smooth functions $u,v$ which satisfy 
\[
\Delta u(\mbr) + k_\star u(\mbr) = 0 \quad \text{and} \quad \Delta v(\mbr) + k v(\mbr) = 0,
\]
over a set $\Omega$ we have that
%
\begin{equation} \label{eqn:Greens 2nd identity}
 (k^2-k_\star^2)\int_\Omega uv\,\dd \mbr
  = \int_\Omega \left( \Delta u v-u\Delta v \right) \,\dd\mbr =  \int_{\d\Omega} \left(\d_\mbnu uv - u\d_\mbnu v\right) \,\dd s(\mbr).
\end{equation}
With $u(\mbr_2)$ substituted for $\ensem{f_{n',N}}(\mbr_2,\lambda_2)$ and $v(\mbr_2)$ substituted for $\mrm U_{n'-n}(k\mbr_1-k\mbr_2)$ we can use the above to deduce:
\begin{equation}
\label{eq:IJ green}
\int_{\Rc_2\setminus D(\mbr_1,a_{12})} \mrm U_{n'-n}(k\mbr_1-k\mbr_2)\ensem{f_{n'}}(\mbr_2,\lambda_2) \dd \mbr_2
=
\frac{\Ic_{n'n}(\mbr_1) - \Jc_{n'n}(\mbr_1) }{k^2-k_\star^2},
\end{equation}
where we defined
\begin{equation}
\begin{aligned}
\label{def:matcal I and J}
    \Ic_{n'n}(\mbr_1) &:= \int_{\d \Rc_2}\mrm U_{n'-n}(k\mbr_1-k\mbr_2)
    \frac{\d \ensem{f_{n'}}(\mbr_2,\lambda_2)}{\d \mbnu_2}
    -
    \frac{\d \mrm U_{n'-n}(k\mbr_1-k\mbr_2)}{\d \mbnu_2}
    \ensem{f_{n'}}(\mbr_2,\lambda_2)\dd A_2,
    \\
    \Jc_{n'n}(\mbr_1) &:= \int_{\d D(\bm{0},a_{12})}\mrm U_{n'-n}(- k\mbr)
    \frac{\d \ensem{f_{n'}}(\mbr + \mbr_1,\lambda_2)}{\d \mbnu}
    -
    \frac{\d\mrm U_{n'-n}(-k\mbr)}{\d \mbnu}
    \ensem{f_{n'}}(\mbr + \mbr_1,\lambda_2)\dd A.
\end{aligned}
\end{equation}
Finally, substituting \eqref{def:matcal K} and \eqref{eq:IJ green} into the governing equation~\eqref{eq:average governing equation} gives
\begin{multline}
\label{eq:averaged governing equation pre-split}
    \ensem{f_{n}}(\mbr_1,\lambda_1) = \mrm T_{n}(\lambda_1) \sum_{n'}\mrm V_{n'-n}(k\mbr_1) g_{n'}
    \\
    +\sum_{n'}\mrm T_n(\lambda_1)\int_\Sc  \left[
    \frac{\Ic_{n'n}(\mbr_1)-\Jc_{n'n}(\mbr_1)}{k^2-k_\star^2}
    + \Kc_{n'n}(\mbr_1)
    \right] \mathfrak{n}(\lambda_2)\dd\lambda_2.
\end{multline}
The above now can be split into two separate equations by noting that the functions $\ensem{f_{n}}(\mbr_1)$, $\Jc_{n'n}(\mbr_1)$ and $\Kc_{n'n}(\mbr_1)$ satisfy the wave equation with wavenumber $k_\star$, while $\mrm V_n(k\mbr_1)$ and $\Ic_{n'n}(\mbr_1)$ satisfy the wave equation with wavenumber $k$. Since solutions of the Helmholtz equation with different wavenumbers are independent, see \cite{gower2021effective} for details, \eqref{eq:averaged governing equation pre-split} can be split into the ensemble wave equation \eqref{eq:ensemble wave equation} containing the terms with wavenumber $k_\star$ and the ensemble boundary conditions \eqref{eq:ensemble boundary equation} containing the terms with wavenumber $k$.

\subsection{The effective eigensystem} \label{sec:effective eigensystem}

Here we deduce a general eigensystem which can be used to determine the effective wavenumber $k_\star$ and write $\ensem{f_{n}}(\mbr_1,\lambda_1)$ in terms of eigenfunctions.

Since  $\ensem{f_{n}}(\mbr_1,\lambda_1)$ satisfies the wave equation \eqref{def:effective wavenumber assumption}, it can be decomposed into the modes
\begin{equation}
\label{eq:fnN decomposition}
\ensem{f_{n}}(\mbr_1,\lambda_1) = \sum_{n_1}F_{nn_1}(\lambda_1)\mrm V_{n_1}(k_\star\mbr_1),
\end{equation}
where $\mrm V_{n_1}$ is defined in \eqref{def:u_n and v_n for a cylinder}. 

The unknowns $k_\star$ and $F_{nn_1}(\lambda_1)$ can be determined by substituting \eqref{eq:fnN decomposition} into \eqref{eq:ensemble wave equation}, which requires the term
\begin{equation}
\begin{aligned}
\label{eq:fn effective modes decomposition after Graf}
\ensem{f_{n}}(\mbr+\mbr_1,\lambda_2)  &= \sum_{n_1n_2}F_{nn_1}(\lambda_2)\mrm V_{n_1-n_2}(k_\star\mbr)\mrm V_{n_2}(k_\star\mbr_1),
\end{aligned}
\end{equation}
where the right side is a result of using Graf's addition theorem (\ref{eq:translation identities}, i) in \eqref{eq:fn effective modes decomposition}.

By substituting \eqref{eq:fn effective modes decomposition after Graf} in $\Kc_{n'n}(\mbr_1)$ \eqref{def:matcal K} we can use the orthogonality of the cylindrical Bessel functions to remove the sum over $n_2$, because only the cases $(n_1-n_2)=(n-n')$ are non-zero. 
Likewise, we can perform the same simplification by substituting \eqref{eq:fn effective modes decomposition after Graf} in $\Jc_{n'n}(\mbr_1)$ \eqref{def:matcal I and J}. The simplifications result in
\begin{equation}
\label{eq:matcal JK regular}
\begin{aligned}
    \Kc_{n'n}(\mbr_1)  
    &=  
   2\pi \mrm W_{n'-n}(k,k_\star) \sum_{n_1}  F_{n'n_1}(\lambda_2)\mrm V_{n_1+n'-n}(k_\star\mbr_1),
    \\
    \Jc_{n'n}(\mbr_1) 
    &=
      - 2\pi\mrm N_{n'-n}(k a_{12},k_\star a_{12})\sum_{n_1} F_{n'n_1}(\lambda_2)\mrm V_{n_1+n'-n}(k_\star\mbr_1),
    \end{aligned}
\end{equation}
where we introduced the notations
\begin{equation} \label{eqn:W and N}
    \begin{aligned}
\mrm W_l(k,k_\star) &:= \int_{a_{12}}^{b_{12}} \mrm H_{l}( kr) \mrm J_{l}(k_\star r)
    \delta\pc(r,\lambda_1,\lambda_2)r\, \dd r ,
\\
\mrm N_l(x, y) &:=
x \mrm H'_{l}(x) \mrm J_l(y) - y\mrm H_{l}(x) \mrm J'_{l}(y).
    \end{aligned}
\end{equation}
%

Finally, substituting \eqref{eq:matcal JK regular} in the ensemble wave equation \eqref{eq:ensemble wave equation}, and again using the orthogonality of the cylindrical Bessel functions, we reach  equation \eqref{eq:regular eigensystem (matrix form)}
%
%
where the following term appears:
\begin{equation}
\Nc_l^{12}(k,k_\star) = 2\pi\frac{\mrm N_l(k a_{12},k_\star a_{12})}{k_\star^2 - k^2} -2\pi\mrm W_l(k,k_\star).
\end{equation}
%
%

\section{Boundary condition for effective waves }
\label{sec:effective boundary}

The eigensystem \eqref{eq:regular eigensystem (matrix form)} is not enough to fully determine the $F_{n n_1}$, to do so we need to  substitute \eqref{eq:fn effective modes decomposition} into the ensemble boundary condition \eqref{eq:ensemble boundary equation}. To achieve this the first step is to use Graf's addition theorem (\ref{eq:translation identities}, iii) with $\mbx=\mbr_1$ and $\mby=-\mbr_2$ to obtain
\begin{equation}
\label{eq:Graf for ensemble boundary condition}
    \mrm U_{n'-n}(k\mbr_1-k\mbr_2)
    =
    \sum_{n_3} \mrm V_{n_3}(k\mbr_1)\mrm U_{n'-n-n_3}(-k\mbr_2)
    =
    \sum_{n_2} \mrm V_{n_2-n+n'}(k\mbr_1)\mrm U_{-n_2}(-k\mbr_2),
\end{equation}
where we used the change of variable $n_2=n+n_3-n'$, and that $|\mbr_1| < |\mbr_2|$ because for $\Ic_{n'n}$ the variable $\mbr_2$ is on the boundary $\partial \Rc_2$, whereas $\mbr_1$ satisfies \eqref{eqn:r1 distance}. Substituting \eqref{eq:fn effective modes decomposition} in $\Ic_{n'n}$ defined by \eqref{def:matcal I and J} and using the above \eqref{eq:Graf for ensemble boundary condition} gives 

\begin{equation}
\label{eqn:Ic regular}
    \Ic_{n'n}(\mbr_1) = \sum_{n_2 n_1} 
 F_{n' n_1}(\lambda_2) \mathcal B_{n_1 n_2} \mrm V_{n_2-n+n'}(k\mbr_1)
\end{equation}
where
\begin{equation} \label{eqn:B integral}
    \mathcal B_{n_1 n_2} = (-1)^{n_2}\int_{\d \Rc_2}\left[\mrm U_{-n_2}(k\mbr_2)
   \frac{\d \mrm V_{n_1}(k_\star\mbr_2)}{\d \mbnu_2}
 - \frac{\d\mrm U_{-n_2}(k\mbr_2)}{\d \mbnu_2}\mrm V_{n_1}(k_\star\mbr_2)
 \right]
  \dd A_2
\end{equation}
then substituting \eqref{eqn:Ic regular} in \eqref{eq:ensemble boundary equation} leads to
\begin{equation}
\label{eq:general ensemble boundary condition-1}
    \sum_{n'} \mrm V_{n'-n}(k\mbr_1) g_{n'}+ \sum_{n'n_2 n_1}\int_\Sc 
 F_{n' n_1}(\lambda_2) \mrm V_{n_2-n+n'}(k\mbr_1)
    \frac{\mathcal B_{n_1 n_2}}{k^2-k_\star^2} \mathfrak{n}(\lambda_2)\,\dd\lambda_2= 0.
\end{equation}
We can further simplify the above by using the orthogonality of the functions $\mrm V_n$ to obtain 
\begin{equation}
\label{eq:general ensemble boundary condition}
    \tcboxmath{
     g_{N}+ \sum_{n' n_1}\int_\Sc 
 F_{n' n_1}(\lambda_2) 
    \frac{\mathcal B_{n_1 (N-n')}}{k^2-k_\star^2} \mathfrak{n}(\lambda_2)\,\dd\lambda_2= 0,
    }
\end{equation}
which holds for every $N$.

When all particles are in a disk, then $\partial \Rc_2$ is a circle and the above simplifies. This is the only case we completely resolve in this paper. Let $R_2$ be the radius of the disk $\Rc_2$, then $n_2 = n_1$ in \eqref{eqn:B integral} which reduces to
\begin{equation}
    \mathcal B_{n_1 n_2} = -2\pi\delta_{n_1  - n_2}N_{n_1}(kR_2,k_\star R_2),
\end{equation}
where $N_{n_1}$ is defined by \eqref{eqn:W and N}${}_2$. Substituting this into \eqref{eq:general ensemble boundary condition} leads to \eqref{eq:circular ensemble boundary condition}.

\section{Elementary proof that thee effective T-matrix is diagonal}
\label{app:T is diagonal}
We provide an elementary proof that $\Tc$ is diagonal when the particles are confined in a disk of radius $\mrm R$. To this end, we consider the scattering  from the modal source $\ui^N$ obtained for $g_n=\delta_{n,N}$: 
The notation $\mathfrak{F}_{n,N}$ is the corresponding $\mathfrak{F}_n$ to the specific incident field with $g_n=\delta_{n,N}$ (compare with \eqref{eq:us compact expression}). 
We then denote by $f_{n,N}^{i}(\sigma)$ the resulting solution of \eqref{eq:f_n^i governing equation} for the specific configuration $\sigma=\mbr_1,\dots,\mbr_{J}$. The rotation by angle $\phi$ of the particles $\mbr_1,\dots,\mbr_{ J}$ correspond to another valid configuration (because the random material is cylindrical), for which the solutions are given by 
\begin{equation}
\label{eq:Foldy-Lax solutions rotated (appendix)}
f_{n,N}^{i}(\mbR_\phi\sigma) = \mrm e^{\mrm i(N-n)\phi} f_{n,N}^i(\sigma)
\end{equation}
This tells us how the rotation of a configuration modifies the coefficient $\mathfrak F_{n,N}$, using \eqref{eq:us compact expression}:
\begin{equation}
\mathfrak F_{n,N}(\mbR_\phi\sigma)
=
\sum_{i=1}^{ J} \sum_{n'=-\infty}^{+\infty}\overline{ \mrm V_{n-n'}(k\mbr_i)} \mrm e^{\mrm i(N-n)\phi} f_{n',N}^i(\sigma).
\end{equation}
Consequently,
\begin{equation}
\ensem{\mathfrak F_{n,N}}
=
\int_\sigma \mathfrak F_{n,N}(\sigma)\proba(\sigma)\,\dd\sigma
=
\int_\sigma \frac{1}{2\pi}\int_0^{2\pi}\mathfrak F_{n,N}(\mbR_\phi\sigma)\proba(\sigma)\,\dd\sigma\dd\phi
=
\delta_{n,N}\int_\sigma \mathfrak F_{n,N}(\sigma)\proba(\sigma)\,\dd\sigma
\end{equation}
Finally, we deduce
\begin{equation}
\tcboxmath{
\Tc_{n,N} = \delta_{n,N}\int \mathfrak F_{n,N}(\mbr_1,\dots,\mbr_{J})\proba(\mbr_1,\dots,\mbr_{J})\,\dd \mbr_1,\dots,\dd\mbr_{J}.}
\end{equation}
This analysis proves that only the diagonal terms of the effective T-matrix are nonzero and can be estimated by $\Tc_{n,n}=\ensem{\mathfrak{F}_{n,n}}$.

\end{appendices}

 \bibliographystyle{plain}
 \bibliography{biblio}

\begin{thebibliography}{10}

\bibitem{bose1973longitudinal}
SK~Bose and AK~Mal.
\newblock Longitudinal shear waves in a fiber-reinforced composite.
\newblock {\em International Journal of Solids and Structures},
  9(9):1075--1085, 1973.

\bibitem{challis2005ultrasound}
RE~Challis, MJW Povey, ML~Mather, and AK~Holmes.
\newblock Ultrasound techniques for characterizing colloidal dispersions.
\newblock {\em Reports on progress in physics}, 68(7):1541, 2005.

\bibitem{chekroun2012time}
Mathieu Chekroun, Lo{\"\i}c Le~Marrec, Bruno Lombard, and Jo{\"e}l Piraux.
\newblock Time-domain numerical simulations of multiple scattering to extract
  elastic effective wavenumbers.
\newblock {\em Waves in Random and Complex media}, 22(3):398--422, 2012.

\bibitem{cotterill2022deeply}
Philip~A Cotterill, David Nigro, and William~J Parnell.
\newblock Deeply subwavelength giant monopole elastodynamic metacluster
  resonators.
\newblock {\em Proceedings of the Royal Society A}, 478(2263):20220026, 2022.

\bibitem{Dubois2011}
J~Dubois, C~Arist{\'{e}}gui, O~Poncelet, and A~L Shuvalov.
\newblock Coherent acoustic response of a screen containing a random
  distribution of scatterers: Comparison between different approaches.
\newblock {\em Journal of Physics: Conference Series}, 269:012004, January
  2011.

\bibitem{foldy1945multiple}
Leslie~L Foldy.
\newblock The multiple scattering of waves. i. general theory of isotropic
  scattering by randomly distributed scatterers.
\newblock {\em Physical review}, 67(3-4):107, 1945.

\bibitem{ganesh2017algorithm}
Mahadevan Ganesh and Stuart~C Hawkins.
\newblock Algorithm 975: Tmatrom—a t-matrix reduced order model software.
\newblock {\em ACM Transactions on Mathematical Software (TOMS)}, 44(1):1--18,
  2017.

\bibitem{gower2019proof}
Artur~L Gower, I~David Abrahams, and William~J Parnell.
\newblock A proof that multiple waves propagate in ensemble-averaged
  particulate materials.
\newblock {\em Proceedings of the Royal Society A}, 475(2229):20190344, 2019.

\bibitem{gower2018characterising}
Artur~L Gower, Robert~M Gower, Jonathan Deakin, William~J Parnell, and I~David
  Abrahams.
\newblock Characterising particulate random media from near-surface
  backscattering: A machine learning approach to predict particle size and
  concentration.
\newblock {\em Europhysics Letters}, 122(5):54001, 2018.

\bibitem{gower2021effective}
Artur~L Gower and Gerhard Kristensson.
\newblock Effective waves for random three-dimensional particulate materials.
\newblock {\em New Journal of Physics}, 23(6):063083, 2021.

\bibitem{gower2019multiple}
Artur~L Gower, William~J Parnell, and I~David Abrahams.
\newblock Multiple waves propagate in random particulate materials.
\newblock {\em SIAM Journal on Applied Mathematics}, 79(6):2569--2592, 2019.

\bibitem{gower2018reflection}
Artur~L Gower, Michael~JA Smith, William~J Parnell, and I~David Abrahams.
\newblock Reflection from a multi-species material and its transmitted
  effective wavenumber.
\newblock {\em Proceedings of the Royal Society A: Mathematical, Physical and
  Engineering Sciences}, 474(2212):20170864, 2018.

\bibitem{gumerov2005computation}
Nail~A Gumerov and Ramani Duraiswami.
\newblock Computation of scattering from clusters of spheres using the fast
  multipole method.
\newblock {\em The Journal of the Acoustical Society of America},
  117(4):1744--1761, 2005.

\bibitem{karnezis2023average}
Aris Karnezis, Paulo~S. Piva, and Art~L. Gower.
\newblock The average transmitted wave in random particulate materials.
\newblock 2023.

\bibitem{koc1998calculation}
S~Koc and Weng~Cho Chew.
\newblock Calculation of acoustical scattering from a cluster of scatterers.
\newblock {\em The Journal of the Acoustical Society of America},
  103(2):721--734, 1998.

\bibitem{kong2004scattering}
Jin~Au Kong, Leung Tsang, Kung-Hau Ding, and Chi~On Ao.
\newblock {\em Scattering of electromagnetic waves: numerical simulations}.
\newblock John Wiley \& Sons, 2004.

\bibitem{lax1952multiple}
Melvin Lax.
\newblock Multiple scattering of waves. ii. the effective field in dense
  systems.
\newblock {\em Physical Review}, 85(4):621, 1952.

\bibitem{linton_multiple_2005}
C.~M. Linton and P.~A. Martin.
\newblock Multiple scattering by random configurations of circular cylinders:
  {Second}-order corrections for the effective wavenumber.
\newblock {\em J. Acoust. Soc. Am.}, 117(6):3413, 2005.

\bibitem{linton2005multiple}
CM~Linton and PA~Martin.
\newblock Multiple scattering by random configurations of circular cylinders:
  Second-order corrections for the effective wavenumber.
\newblock {\em The Journal of the Acoustical Society of America},
  117(6):3413--3423, 2005.

\bibitem{martin2003acoustic}
Paul~A Martin.
\newblock Acoustic scattering by inhomogeneous obstacles.
\newblock {\em SIAM Journal on Applied Mathematics}, 64(1):297--308, 2003.

\bibitem{martin2006multiple}
Paul~A Martin.
\newblock {\em Multiple scattering: interaction of time-harmonic waves with N
  obstacles}.
\newblock Number 107. Cambridge University Press, 2006.

\bibitem{mishchenko2006multiple}
Michael~I Mishchenko, Larry~D Travis, and Andrew~A Lacis.
\newblock {\em Multiple scattering of light by particles: radiative transfer
  and coherent backscattering}.
\newblock Cambridge University Press, 2006.

\bibitem{EffectiveTMatrix}
Kevish~K. Napal.
\newblock Julia package {EffectiveTMatrix.jl}, August 2023.

\bibitem{ni2010achieving}
Yaxian Ni, Lei Gao, and Cheng-Wei Qiu.
\newblock Achieving invisibility of homogeneous cylindrically anisotropic
  cylinders.
\newblock {\em Plasmonics}, 5:251--258, 2010.

\bibitem{pendry2004reversing}
John~B Pendry and David~R Smith.
\newblock Reversing light with negative refraction.
\newblock {\em Physics today}, 57(6):37--43, 2004.

\bibitem{rohfritsch2019numerical}
Adrien Rohfritsch, Jean-Marc Conoir, R{\'e}gis Marchiano, and Tony
  Valier-Brasier.
\newblock Numerical simulation of two-dimensional multiple scattering of sound
  by a large number of circular cylinders.
\newblock {\em The Journal of the Acoustical Society of America},
  145(6):3320--3329, 2019.

\bibitem{sheng2007introduction}
Ping Sheng and Bart van Tiggelen.
\newblock Introduction to wave scattering, localization and mesoscopic
  phenomena., 2007.

\bibitem{smith2022tailored}
Michael~JA Smith and I~David Abrahams.
\newblock Tailored acoustic metamaterials. part i. thin-and thick-walled
  helmholtz resonator arrays.
\newblock {\em Proceedings of the Royal Society A}, 478(2262):20220124, 2022.

\bibitem{smith2022asymptotics}
MJA Smith, PA~Cotterill, D~Nigro, WJ~Parnell, and ID~Abrahams.
\newblock Asymptotics of the meta-atom: plane wave scattering by a single
  helmholtz resonator.
\newblock {\em Philosophical Transactions of the Royal Society A},
  380(2237):20210383, 2022.

\bibitem{Tishkovets2011}
Victor~P. Tishkovets, Elena~V. Petrova, and Michael~I. Mishchenko.
\newblock Scattering of electromagnetic waves by ensembles of particles and
  discrete random media.
\newblock {\em Journal of Quantitative Spectroscopy and Radiative Transfer},
  112(13):2095--2127, September 2011.

\bibitem{torrent2006homogenization}
Daniel Torrent, Andreas H{\aa}kansson, Francisco Cervera, and Jos{\'e}
  S{\'a}nchez-Dehesa.
\newblock Homogenization of two-dimensional clusters of rigid rods in air.
\newblock {\em Physical review letters}, 96(20):204302, 2006.

\bibitem{Tsang1982}
L.~Tsang, J.~A. Kong, and T.~Habashy.
\newblock Multiple scattering of acoustic waves by random distribution of
  discrete spherical scatterers with the quasicrystalline and
  {P}ercus{\textendash}{Y}evick approximation.
\newblock {\em The Journal of the Acoustical Society of America},
  71(3):552--558, March 1982.

\bibitem{twersky1962scatteringI}
Victor Twersky.
\newblock On scattering of waves by random distributions. i. free-space
  scatterer formalism.
\newblock {\em Journal of Mathematical Physics}, 3(4):700--715, 1962.

\bibitem{twersky1962scatteringII}
Victor Twersky.
\newblock On scattering of waves by random distributions. ii. two-space
  scatterer formalism.
\newblock {\em Journal of Mathematical Physics}, 3(4):724--734, 1962.

\bibitem{Varadan1983}
V.~K. Varadan, V.~N. Bringi, V.~V. Varadan, and A.~Ishimaru.
\newblock Multiple scattering theory for waves in discrete random media and
  comparison with experiments.
\newblock {\em Radio Science}, 18(3):321--327, May 1983.

\bibitem{vynck2021light}
Kevin Vynck, Romain Pierrat, R{\'e}mi Carminati, Luis~S Froufe-P{\'e}rez, Frank
  Scheffold, Riccardo Sapienza, Silvia Vignolini, and Juan~Jos{\'e} S{\'a}enz.
\newblock Light in correlated disordered media.
\newblock {\em arXiv preprint arXiv:2106.13892}, 2021.

\bibitem{waterman1961multiple}
Peter~Cary Waterman and Rohn Truell.
\newblock Multiple scattering of waves.
\newblock {\em Journal of mathematical physics}, 2(4):512--537, 1961.

\end{thebibliography}

\end{document}